%% file: article.tex
\documentclass[aps,prf,groupedaddress]{revtex4-2}

\usepackage{comment}
\usepackage{siunitx}

\usepackage{graphicx}
\usepackage{dcolumn}
\usepackage{bm}
\usepackage{adjustbox}

\usepackage[utf8]{inputenc}
\usepackage[T1]{fontenc}
\usepackage{mathptmx}
\usepackage{etoolbox}
\usepackage{xcolor}
\usepackage{array, multirow, boldline}

\graphicspath{ {./plot/} {./FIGS/} {./BERENGERE/} }

\makeatletter
\def\@email#1#2{%
 \endgroup
 \patchcmd{\titleblock@produce}
  {\frontmatter@RRAPformat}
  {\frontmatter@RRAPformat{\produce@RRAP{*#1\href{mailto:#2}{#2}}}\frontmatter@RRAPformat}
  {}{}
}%
\makeatother

\begin{document}

\newcommand{\berengere}[1]{{\color{black} #1}}
\newcommand{\caro}[1]{{\color{black} #1}}

\title{Study of the wall pressure variations on the stall inception of a thick cambered profile at high Reynolds number}

\author{Caroline Braud}
\affiliation{Nantes Université, CNRS, Centrale Nantes, Laboratoire LHEEA, 44300, Nantes, France}
\affiliation{CSTB, Centre Scientific et Technique du Bâtiment, Nantes, France}

\author{B\'ereng\`ere Podvin}
\affiliation{Université Paris-Saclay, CNRS, CentraleSupélec, Laboratoire EM2C, 91190, Gif-sur-Yvette, France.}

\author{Julien Deparday}
\affiliation{IET, OST - Eastern Switzerland University of Applied Sciences, Rapperswil, Switzerland}


\begin{abstract}
We present an experimental study of the aerodynamic forces on a thick and cambered airfoil at a high Reynolds number (\num{3.6e6}), which is of direct relevance for wind turbine design.
Unlike thin airfoils at low chord-based Reynolds numbers, no consistent description currently exists for the stall process on such airfoils.
We consider two chord-wise rows of instantaneous wall pressure measurements, taken simultaneously at two spanwise locations over a range of angles of attack.
\berengere{ We show that around maximum lift conditions,  a strong asymmetry is observed in the statistics of the normal force on each chord.}
In this range of angles of attacks, 
the pressure fluctuations are largest in the adverse pressure gradient region, and the  fluctuation peak along the chord is systematically located directly upstream of the mean steady separation point, indicating  intermittent flow separation. 
Moreover, the fluctuations are characterized by bi-stability in both space and time: for each spanwise location, large excursions of the local wall pressure between two different levels can be observed in time (jumps), and these excursions are highly anti-correlated between the two spanwise locations (spatial switches). The characteristic time scale for the switches is found to be well correlated with the amplitudes of the fluctuations. 
Application of Proper Orthogonal Decomposition (POD) analysis to each row of sensors confirms that the flow separation is an inherently local, three-dimensional and unsteady process that occurs in a continuous manner when the angle of attack increases.
The correlation between the dominant POD mode amplitudes is found to be a good indicator of bi-stability.
\berengere{ For all angles of attack}, most of the fluctuations can be captured with the two most energetic POD modes. 
This suggests that force fluctuations near the maximal lift could be modelled by a low-order approach, for monitoring and control purposes.

\end{abstract}
\maketitle

\input{intro.tex}

\input{set-up.tex} 
\input{resultsJ5.tex}
\bibliographystyle{plainnat}
\bibliography{biblio.bib}

\end{document}

%% file: intro.tex
\section{Introduction}

Understanding the flow over airfoils is crucial for numerous applications such as aircraft or wind turbines.
These systems evolve in the atmosphere, thus with strong interactions with the turbulence from atmospheric boundary layer flows \citep{kaimal1994}.
Turbulence modifies the flow structure and dynamics over aerodynamic surfaces and leads to additional fatigue or loss of controllability.
Many recent studies have demonstrated the importance of such phenomena on 2D airfoils submitted to different turbulent inflows in controlled environments, i.e. wind tunnels \citep{devinant2002,sicot2006,breuer2018,mishra2022}. 
Most of the turbulent inflow effects inducing load variations occur for angles of attack close to the maximum lift value, at which, for example, wind turbines operate under optimal conditions, or during the take-off or landing of an aircraft.
Understanding the flow physics at these angles of attack however remains challenging, even in static conditions. 
The flow of an airfoil undergoing static stall and the resulting aerodynamic force are unsteady, and the dynamic features of static stall are qualitatively similar to those of  the dynamic stall \citep{lefouest_dynamics_2021}.
Understanding better static stall physics may help to improve further the dynamic stall phenomena which have been investigated by numerous studies (see e.g. \citep{McCroskey1982, Eldredge2019, Visbal2017, Chiereghin2020}).

The description of the flow physics on airfoils is generally based on a 2D scenario with 3 main regions of interest (1) a boundary layer developing on the upper surface up to (2) the separation point where the flows starts to detach which then interacts with (3) the wake at the trailing edge of the airfoil.
Each of these regions is characterized by complex phenomena, which depend on many parameters and in particular the Reynolds number.

\berengere{ Due to the airfoil geometry and angle of attack, the flow over the airfoil is subject to a continuously  varying spatial
pressure gradient, and the boundary layer is therefore not in  the  equilibrium state that can be  expected
for instance for a fully developed turbulent boundary layer on a flat plate. }
The boundary layer state depends on the surface roughness, the turbulence intensity and the Reynolds number, the effects of which are difficult to disentangle \citep{kerho1997,wang2014,mishra2022}.
The flow can remain laminar within the entire recovery region at low chord-based Reynolds number (smaller than \num{3e4}) if the surface roughness, the turbulence intensity and the angles of attacks are low.
With the increase of the Reynolds number, the transition to a turbulent boundary layer moves towards the leading edge, and can result in the formation of a laminar separation bubble.
At Reynolds numbers larger than \num{e5}, the length of the separation bubble is generally of the order of a few percent of the airfoil chord and thus does not greatly alter the pressure from its normal attached distribution \citep{lissaman1983}.

The airfoil performance is also greatly affected by the flow separation over the airfoil which depends on the state of the boundary layer, the adverse pressure gradient and the wake.
The flow separation position moves along the chord, from which a shear layer is formed at the interface between the low-speed separated flow region and the free stream.
Once the flow is separated, the shear layer vortices may influence wake vortex shedding characteristics.
Wake vortices are formed in the near wake region and are shed alternately on the upper and lower surface of the airfoil. 
They are present at all angles of attack, as observed by \citet{Yarusevych2006}.
\citet{yarusevych09} showed that universal scaling based on local length scales could be identified in both the separated shear layer and the wake.
Furthermore, the wake can be influenced by the three-dimensionality of the flow field \citep{neal2023three}.
{\berengere Despite a possible spanwise dependence of the flow dynamics,
for load evaluation  statistical invariance is generally assumed in the spanwise direction at all angles of attack,  
and measurements are only taken along a single chord..}

When the flow is completely separated over the airfoil, the lift force decreases, which is often referred to as the stall phenomenon. 
\citet{gault1957} found that the type of stall observed was highly dependent on the airfoil geometry and on the Reynolds number. 
For different thickness ratios ranging from \num{0} to \num{24}\% of the chord, and Reynolds numbers ranging from \num{0.7} to \num{25e6}, Gault was able to distinguish four types of stall.
The leading-edge and the thin-airfoil stalls mostly happen on symmetric, thin airfoils and at low chord-based Reynolds numbers. 
It is abrupt and fast, occurring directly at the leading edge.
The trailing-edge stall is a progressive displacement of the boundary layer flow separation from the trailing-edge to the leading-edge, as described for instance by \citet{soulier2021} using flow field measurements.
The last type of stall is a combined trailing-edge and leading-edge stall which was observed by \citet{bak1999} using a moderately thick airfoil and high chord-based Reynolds number experiments, $Re_c=\num{1.3e6}$. 
Moderately thick airfoils are widely used for engineering systems operating in the atmosphere such as wind turbines blades, as they yield less load fluctuations near the maximum lift angle of attack when large inflow variations are encountered compared to thin airfoils.
The flow physics have been widely investigated for symmetric shapes and at low chord-based Reynolds numbers (see e.g. \citep{lissaman1983,yarusevych09,yarusevych09}), but fewer studies exist at high Reynolds numbers above \num{e6}, and/or for (moderately-) thick ($>$ 20\% of the chord) and asymmetric shapes.

Understanding the flow physics and stall behavior at these high Reynolds numbers remains a challenge for both experimental and numerical approaches.
Although transitional regimes $Re_c\simeq$ \num{2.5e4} - \num{e5} have been studied in detail with direct (DNS) or large eddy (LES) simulations \citep{alMutairi2017,elJack2020}, the cost of such approaches still remains prohibitively high at these Reynolds numbers, despite efforts to relax LES resolution requirements \citep{tamaki20}.   
The URANS formalism then represents a viable alternative, and has been successful to predict hysteresis \citep{mittal2002,mittal2014}, as well as 3D-effects \citep{bertagnolio2005,richez2015,plante2020}, but URANS prediction of separation near or at stall should be viewed with caution.
Bifurcation analysis of the stall at high Reynolds numbers has provided useful insight but has been so far limited to a 2D approach \citep{busquet21}.

{\berengere Investigations at high Reynolds numbers also represent an experimental challenge.}
Small facilities rapidly experience disproportionate compressibility effects to reach high Reynolds numbers. 
Adequate wind tunnels for this purpose are generally either of very large size, such as those typically used in industry, or pressurised facilities that were specifically designed to increase the Reynolds number \citep{Miller2019, brunner_al2021}.
Experiments are thus more directed towards time-averaged, global force measurements rather than local and unsteady characterisations.
{\berengere Table \ref{tab:bladeshape} presents a summary of different
studies performed at high Reynolds numbers for different airfoil thicknesses. Except in one case,
all stalls are of the trailing edge type.} 
Furthermore, at high Reynolds numbers, near the maximum lift, the flow becomes three-dimensional and displays complex dynamics, while measurement tools remain limited to study these dynamics, especially for large scale facilities.
\citet{manolesos2014,ragni2016} have analyzed three-dimensional flow separation in wind tunnel experiments at chord-based Reynolds numbers of $O(10^6)$. 

{\berengere Flow visualizations provided evidence of a spanwise distribution of stall cells near the maximum lift value. } 
For similar Reynolds numbers, \citet{olsen2020} have recently shown the high sensitivity of the lift curve close to the maximum lift to small differences in the experimental execution. 
\citet{brunner_al2021} have investigated the Reynolds number effect ranging from \num{5e5} to \num{7.9e6} of a moderately thick airfoil.
Fundamental change of flow behavior was observed around $Re_c=\num{2e6}$: the stall gradually shifts from trailing-edge stall to leading-edge stall. 
{\berengere In contrast,  only trailing- edge stalls were found 
by \citet{neunaber2022} when investigating a 2-D blade of similar thickness  but of different shape. 
The   blade} was extracted from a 2MW turbine at 80\% of the rotor diameter (20\% thick)
and was studied {\berengere at a chord-based Reynolds number of \num{4.7e6}}. 
The same blade section has {\berengere also} been investigated {\berengere in} a lower Reynolds number range $O(10^5)$ \citep{mishra2022}. 
Both Reynolds number ranges exhibit a trailing-edge stall, contrary to what was found for a different airfoil shape by \citet{brunner_al2021}.

Moreover, \citet{neunaber2022} found that the normal forces measured at two different spanwise locations alternated between two quasi-steady states, thus pointing out to bi-stability in space and time. 
This bi-stability was correlated with the unsteady displacement of the separation point.
These recent findings highlight the fact that the stall process at high Reynolds number is a complex phenomenon, still not fully understood.

Using the same configuration as \citet{neunaber2022}, the present paper focuses on investigating the flow bi-stability evolution with the increase of the angle of attack. 
{\berengere  Comparison of the two chords 
provides a check on the statistical bi-dimensionality of the flow, which is often assumed a priori.
In addition, the spatial organization of the fluctuations is described using} 
Proper Orthogonal Decomposition (POD), which has proved to be a useful tool to extract coherent structures from pressure measurements in a variety of turbulent flows, for such phenomena as jet mixing layer instabilities \citep{arndt_long_glauser_1997}, flow separation over a blunt flat plate \citep{tenaud16}, or flapping instability of a sail \citep{deparday18}.
POD has also been applied to airfoils at moderate Reynolds numbers ($O(10^5)$): \citet{villegas16} have investigated the relationship between vortex shedding and aerodynamic force fluctuations, while \citet{ribeiro17} and \citet{yang21} have focused on the generation mechanisms of airfoil tonal noise. 
In this study, the high Reynolds number considered is reached thanks to the large scale test section of the CSTB climatic wind tunnel, i.e. a test section of size \SI{5}{\metre} by \SI{6}{\metre}. 
We first present in section \ref{ssec:setup} the set-up of experiments and we briefly review the POD method. 
We then give a general statistical description of the normal force in section \ref{sec:statistics}.
Section \ref{sec:local} is focused on the spatio-temporal characteristics of the flow bi-stability.
POD analysis of the wall pressure fluctuations is provided in section \ref{sec:pod}. 
The main spatio-temporal features of the pressure fluctuations, and their dependence on the angle of attack, are summarized in section \ref{sec:conclusions}.

\begin{table}[htbp] 
\centering
\begin{tabular}{ccccc}
{\bf Paper Reference} & {\bf Airfoil name} & {\bf Thickness} & {\bf $Re_c$} & {\bf Type of stall according to Gault \cite{gault1957}}  \\
Present study & Scanned & $20\%$c & \num{3.4e6}  & TE\\ 
\cite{brunner_al2021} & NACA0021 &  $21\%$c & $\num{5e5} \leq Re_c \leq \num{7.9e6}$ & from TE to LE  \\
\cite{manolesos2014} & NC &$18\%$c & \num{e6} & TE\\ 
\cite{ragni2016} & NACA64-418 & $18\%$c &  \num{e6} & TE\\
\cite{olsen2020} & NACA63-418 & $18\%$c & \num{3.4e6} & TE\\
\cite{neunaber2022} & Scanned & $20\%$c & \num{4.7e6} & TE
\end{tabular}
\caption{Blade characteristics of experiments at high Reynolds numbers using thick profiles (reference [30] to [34]). NC means Not Communicated, Scanned means it has been scanned from a 2MW wind turbine blade.}
\label{tab:bladeshape}
\end{table}

%% file: set-up.tex
\section{Experimental set-up and methods}
\label{ssec:setup}

\subsection{The CSTB Wind tunnel}

\begin{figure}
    \includegraphics[width=0.9\linewidth]{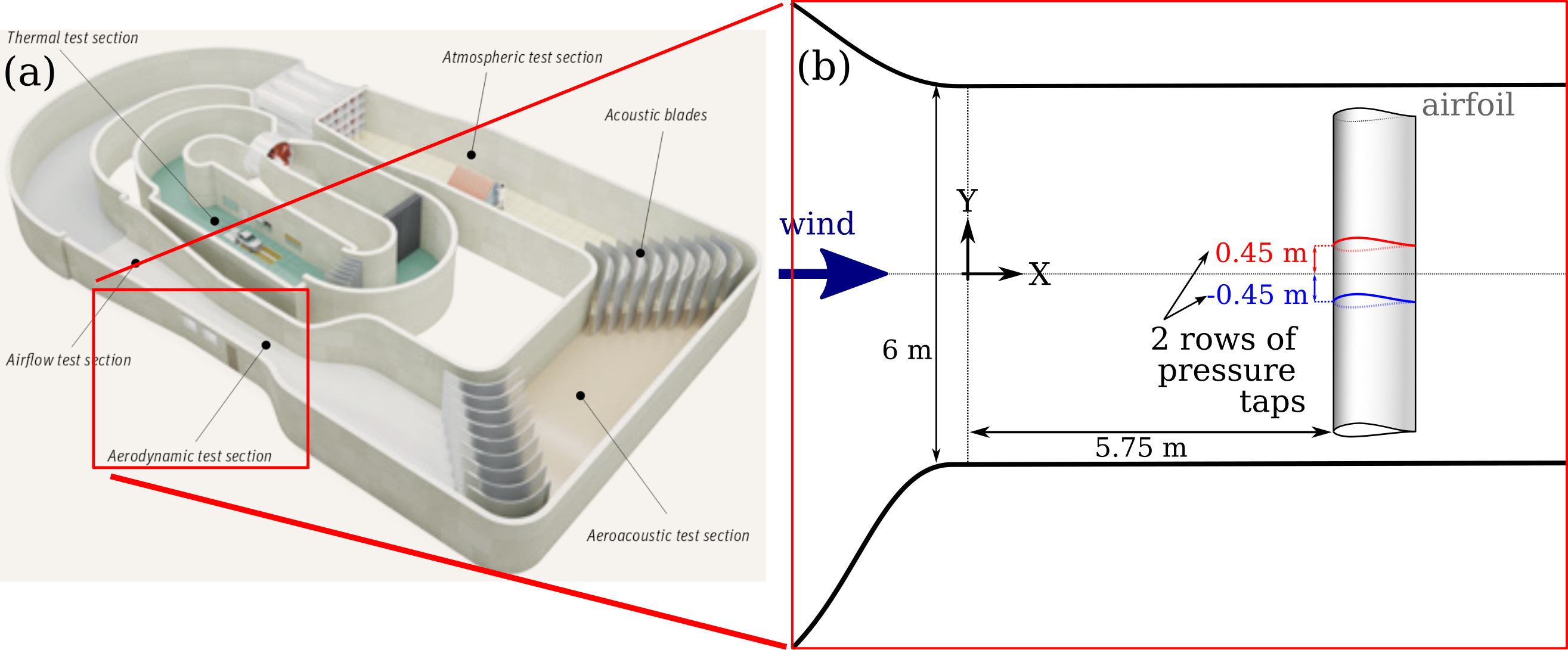}
	\caption{a) The CSTB wind tunnel, b) schematics of the 2D blade installation in the aerodynamic test section with the two rows of 78 pressure taps each in the chord wise direction at two different spanwise locations (red and blue corresponds to respectively $Y^{+}$ and $Y^{-}$ locations).
\berengere{  c) The inflow is measured 2.65 m upstream of the profile by a cobra probe.}
From \citet{neunaber2022}.}
	\label{fig:Setup}
\end{figure}

Experiments have been performed in the aerodynamic test section of the CSTB Wind tunnel (figure \ref{fig:Setup}a.). 
The test section is \SI{6}{\metre} wide and \SI{5}{\metre} high and has a length of \SI{12}{\metre}.
The free stream wind can reach a speed of \SI{70}{\metre\per\second} and is controlled to keep a constant and uniform velocity, taking into account the air density variations.
The turbulence intensity is less than 1.5\% in the empty test section.

\subsection{2D blade section}
\label{ssec:profile}

The 2D blade has a chord of \SI{1.25}{\metre} and a span of \SI{5}{\metre}, which is one meter smaller than the test section, so that wall boundary layers of the test section (\SI{20}{\centi\metre} thick at maximum), do not interact with the blade.
\berengere{ The resulting blockage is thus $8\%$ at \ang{24}, which compares well with other studies carried out at high Reynolds 
numbers \cite{bak2022}. 
The blockage effect is likely to have some influence on the effective angle of attack.
Computations at zero incidence using  correction laws provided by  \cite{barlow1999}   yielded an effective
angle of attack of 1 degree, although this result should be interpreted with caution near flow separation owing to the limitations of these estimation methods
which are based on potential flow theory. 
The aspect ratio of the blade is 4, which is one of the largest ratio set in the literature at large Reynolds numbers (see e.g. \cite{bak2022}).
To prevent large end-plates from increasing the blockage, the extremities of the wing are set free.
The blade is not spanning the entire test section so that tip vortices are formed on both sides of the model.
It was verified that the tip vortices do not interact with the central part of the wing by using nylon tufts installed on the airfoil surface prior to the experiments.}

The blade has been set at \SI{5.75}{\metre} from the inlet on a lattice support structure that allows to set pre-defined angles of attack ($AoA$).
\berengere{ To obtain a good yaw alignment of the blade with the inflow, a cross-line laser level is used to align a transverse line marked at 10\% of the chord perpendicular to the build-in line on the wind tunnel floor which is parallel to the incoming wind direction.
The yaw angle is therefore zero with an accuracy of $\pm$ \ang{0.1}.}

The blade profile comes from 3D scans of a commercial \SI{2}{\mega\watt} wind turbine blade at $80\%$ of its rotor diameter. 
It can be approximated by a NACA63-3-620 airfoil shape with a modified camber (4\% instead of 2\%). 
The maximum thickness is located at around 33\% of the chord while the maximum camber is located at approximately 49\% of the chord. 
Following \citet{gault1957}, we expect to find either trailing edge stall or combined trailing edge and leading edge stall because of the thick and highly cambered profile shape.

\subsection{Unsteady wall pressure measurements}

The 2D blade section was equipped with two rows of 78 unsteady wall pressure taps each in the chord-wise direction, located at two spanwise positions equally distributed from the mid-span of the blade ($Y^{-} = \SI{-450}{\milli\metre}$ and $Y^{+} = \SI{+450}{\milli\metre}$) (figure \ref{fig:Setup}(b)). 
The chord-wise spacing between the pressure taps is \num{0.026}$c$, where $c$ is the chord of the airfoil. 
The pressure taps were connected to five multiplexed EPS pressure scanners of 32 channels each, using \SI{1.5}{\metre} vinyl tubes with an internal diameter of \SI{0.8}{\milli\metre}.
Two ranges of the pressure sensors were used depending on their location, \SIrange{0}{7}{\kilo\pascal} near the leading edge suction peak and \SIrange{0}{2.5}{\kilo\pascal} elsewhere, with a precision of $\pm0.03\%$ of the full measurement range. 
The transfer function of the whole system (tubes plus sensor cavity) has been measured off-line at a sampling frequency of \SI{1024}{\hertz} following the methodology of \citet{holmes1986} and \citet{withmore1996} and has been taken into account in the processing of the pressure data. 
The signal acquisition was performed using two National Instrument acquisition boards linked by real-time system integration for synchronization purposes. 
During the measurements, the sampling frequency was \SI{200}{\hertz}. 
Measurements were carried out at eight angles of attack between \SIrange{6}{24}{\degree}. 
The measurement duration was \SI{120}{\second} for each angle of attack. 
The inflow velocity was $U_0=$ \SI{40.5}{\metre\per\second} which leads to a chord-based Reynolds number of $Re_c = $ \num{3.4e6}. 

In this study, the mean normal force coefficient $C_N$ and the mean pressure coefficient $C_p = \Delta p/q_0$ (with $\Delta p$ the differential pressure between wall pressure and the dynamic pressure $q_0$) were primarily investigated. 
The normal force coefficient indicates the normal force acting on the airfoil in the airfoil coordinate system, and was obtained by integrating the pressure coefficient around the airfoil. 
As the lift force is calculated from the normal and tangential forces, and the latter are only partially retrievable from the pressure coefficient (the viscous part of the tangential force was ignored using this measurement method), we focus on the normal force coefficient in this study. 

\subsection{Detection of the flow separation}
\label{sec:separation_point}

For the analysis of the wall pressure variations presented in this paper, the region of the flow separation is a salient feature and must be correctly detected and therefore correctly defined.
However, unlike the laminar case, turbulent separation is characterized by "a spectrum of states", as first pointed out by \citet{kline57}.
\citet{sandborn61} suggested that separation should be viewed as a transition process over a region of variable length, and proposed a model with two distinct separation regions, corresponding to intermittent and steady separation.
The model was validated by \citet{sandbornliu68} and extended by \citet{simpson81}, who introduced several new definitions, based on the amount of time reverse flow is observed at a particular location. 
In particular, the following regions were defined: i) incipient detachment, for which reverse flow is observed 1\% of the time, ii) intermittent transitory detachment, corresponding to 20\% of the time and iii) transitory detachment, corresponding to 50\% of the time. 
\citet{simpson81} found that transitory detachment coincided with a zero value for the time-averaged wall shear stress. 

\citet{sandbornliu68} used Stratford's criterion \citep{stratford59} theoretically developed using simplified turbulent equations to identify the steady separation point. 
An extension of this criterion was used by \citep{celik22, neunaber2022}, where the separation point was defined as the first location where the pressure gradient becomes smaller than a given threshold.
An advantage of this definition is that it can be applied to instantaneous pressure coefficients, making it possible to track the separation point in time. 
In addition to this separation point definition, we will use another definition for the intermittent separation region, based on the location of the pressure fluctuation peak. 
This region is located upstream of the steady separation point as already pointed out by \cite{sandbornliu68,kline57,simpson81}. 
The location of the fluctuation peak maximum, which is defined only statistically, will be referred to throughout the paper as the {\it intermittent separation point.}

\subsection{Proper orthogonal decomposition}

To obtain further insight into the organization of the fluctuations, we apply Proper Orthogonal Decomposition (POD).
POD \cite{lumleyPOD} is a statistical technique that provides a spectral decomposition of the covariance of the fluctuations, and makes it possible to represent the spatio-temporal pressure signal $p$ as a superposition of spatial patterns $\Phi_n(x)$, the amplitude of which varies in time.
The reader is referred to \cite{HLB} for more details.
Here we will consider the fluctuating pressure signal for each spanwise location and apply POD independently to $Y^+$ and $Y^-$. 
We checked that excluding the pressure side from the decomposition did not alter the shape of the modes and the domain of analysis was thus restricted to the suction side. 
Given $N_T$ time measurements at times $t_k$ of the pressure signal $p(x_i,t_k)$ taken at $N$ different chord positions $x_i$, one builds the autocorrelation matrix $M$
\[ M_{ij} = \frac{1}{N_T} \sum_{k=1}^{N_T} p(x_i,t_k) p(x_j,t_k) \] from which one extracts $N$ eigenvalues $\lambda_n$ and $N$ eigenvectors or empirical modes $\Phi_n(x)$.
We note that the normalized eigenmodes $\Phi_n$ are defined within an arbitrary sign change.
The $\lambda_n$ can be ordered following $\lambda_1 \ge \lambda_2 \ge \ldots \lambda_N$ and represent the contributions of each mode $n$ to the total variance.
In addition, the instantaneous pressure fluctuation at any location can then be reconstructed as
\begin{equation}
p(x_i,t_k)=\sum_{n=1}^{N} \lambda_n^{1/2}  a_n(t_k) \Phi_n(x_i)
\end{equation}
and the evolution of each normalized mode amplitude $a_n$ can be considered independently. 

%% file: resultsJ5.tex
\section{Statistical description} %
\label{sec:statistics}

\subsection{Normal force}

\begin{figure}[htbp]
\centerline{\includegraphics[width=0.6\textwidth]{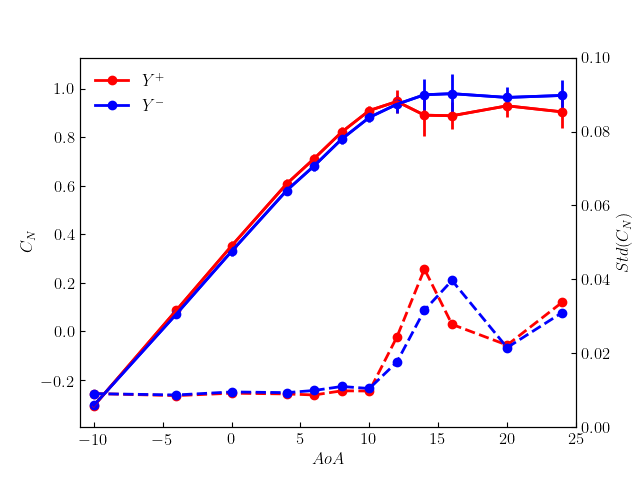}}
\caption{ Normal force coefficient $C_N$ in solid lines and standard deviation of $C_N$ in dashed lines from both spanwise locations, $Y^+$ and $Y^- $.}

\label{fig:CN}
\end{figure}

The normal force obtained by integration of the chord-wise pressure distribution at both spanwise locations, $Y^+$ and $Y^- $, shows a typical lift distribution of wind turbine blade sections (see figure \ref{fig:CN}).
At low angles of attacks ($\leq \ang{8}$), the normal force increases linearly with the angle of attack.
Then a transition occurs with a significant but progressive modification of the slope, becoming almost null above an angle of attack of \ang{12}. 
This slope evolution corresponds to a boundary layer flow separation at the trailing edge, as is confirmed from the trailing edge flattening of the mean pressure chord-wise distribution (see figure \ref{fig:CPmean}).
The evolution of the normal force for both rows of pressure $Y^+$ and $Y^- $ is almost identical until the angle of attack \ang{12}, which proves a good symmetry of the flow upstream and over the airfoil.
At higher angles of attack, on the plateau of the maximum normal force, the row $Y^+$ (in red in figure~\ref{fig:CN}) follows the same trend as $Y^-$ but with values approximately 5\% smaller.\\

\begin{figure}[htbp]
\centerline{\includegraphics[width=0.5\textwidth]{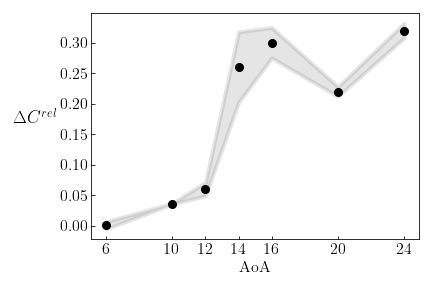} }
\caption{Asymmetry coefficient $\Delta C^{rel} = 2 (C_N(Y^+)-C_N(Y^-))/(C_N(Y^+)+C_N(Y^+))$  for the complete data set (solid points). Grey areas represent the results when only 25\% of the data set length are used.}
\label{asym}
\end{figure}

\berengere{ The symmetry observed when the flow is fully attached breaks at higher angles of attack.
The asymmetry coefficient  $\Delta C^{rel}= 2 (C_N(Y^+)-C_N(Y^-))/(C_N(Y^+)+C_N(Y^+))$ is represented in figure~\ref{asym}.
The shaded area of the curve delimits the values obtained when segments corresponding to 25\% of the signal length were used.
The narrow grey area confirms that the presence of asymmetry is a robust feature of the flow.
Two chords of measurements are naturally not enough to characterize in detail the structure of the flow in the spanwise direction, however they give information about the loss of two-dimensionality of the flow statistics, which is usually taken for granted.
We note that symmetry-breaking behavior has been observed in other symmetric geometries such as the wake of a sphere (\cite{kn:ormieres99}, \cite{kn:fabre08} as well as
for the flow behind an Ahmed body \cite{kn:evstafyeva17}).}

The standard deviations, reported in dashed lines in figure \ref{fig:CN}, start to increase at \ang{10} and then progressively increase with the angle of attack, with a maximum at \ang{14} for $Y^+$ and at \ang{16} for $Y^{-}$, followed by a small decrease at \ang{20}. 
This indicates strong lift fluctuations at the maximum lift that will be analysed further in 
the following sections.

\subsection{Chord-wise wall pressure distribution}
\label{subsec:pressureDist}

\begin{figure}[htbp]
\begin{tabular}{c}
\includegraphics[width=0.6\textwidth]{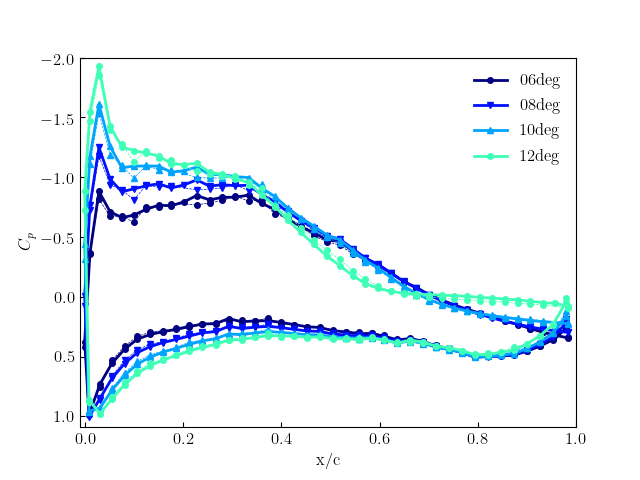} \\
\includegraphics[width=0.6\textwidth]{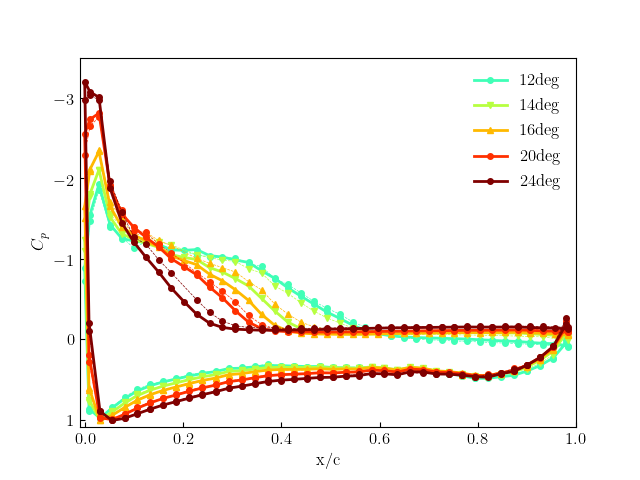}  \\
\end{tabular}
\caption{Pressure coefficient ($C_p$) distribution in chordwise direction $x/c$ for both spanwise locations ($Y^+$ in dotted lines and $Y^-$ in solid lines) for different angles of attack: a) from \ang{6} to \ang{12}, b) from \ang{12} to \ang{24}. The markers indicate the pressure measurement locations.}
\label{fig:CPmean}
\end{figure}

\begin{figure}[htbp]
\begin{tabular}{c}
\includegraphics[width=0.7\textwidth]{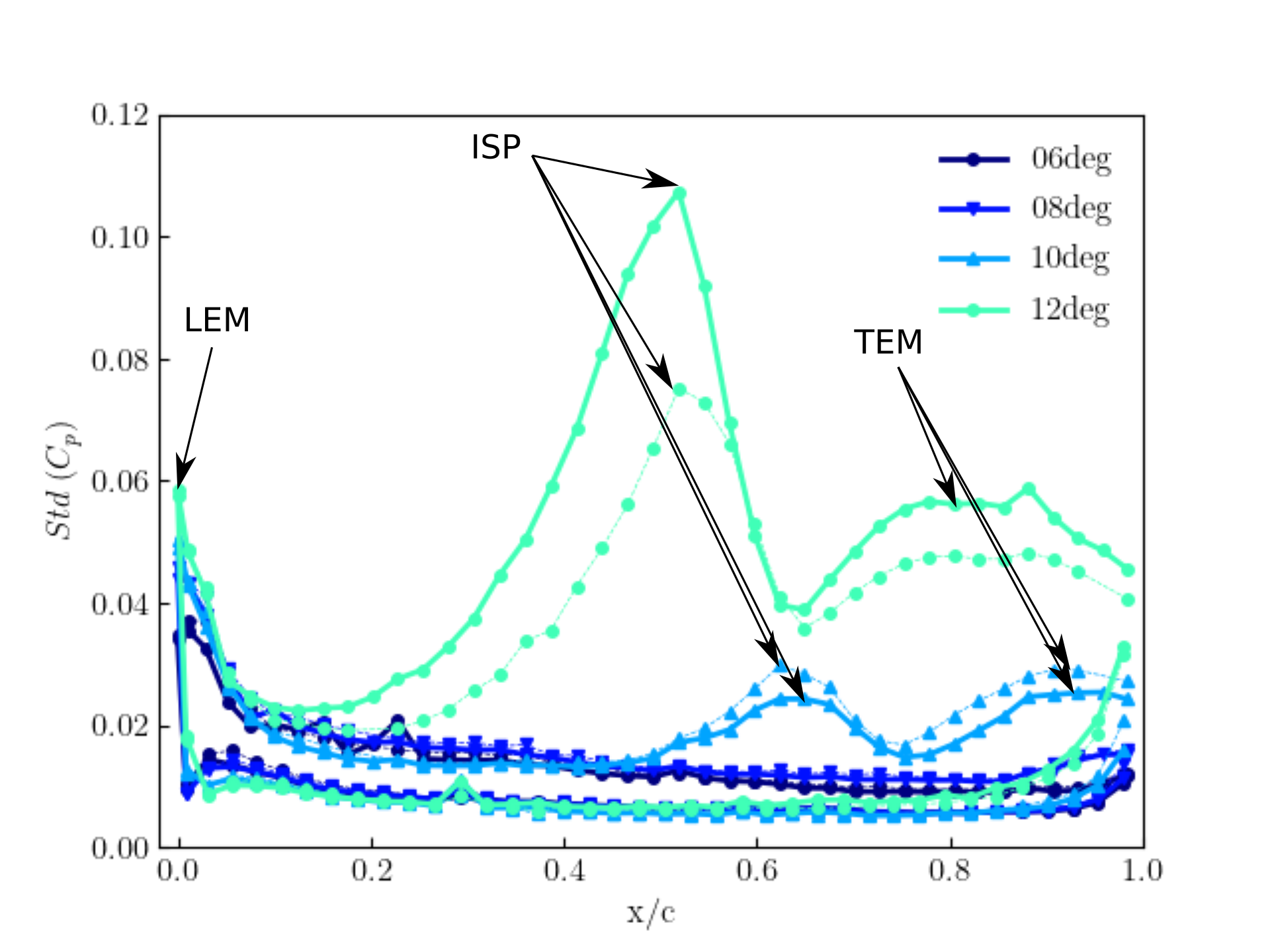}  \\
\includegraphics[width=0.7\textwidth]{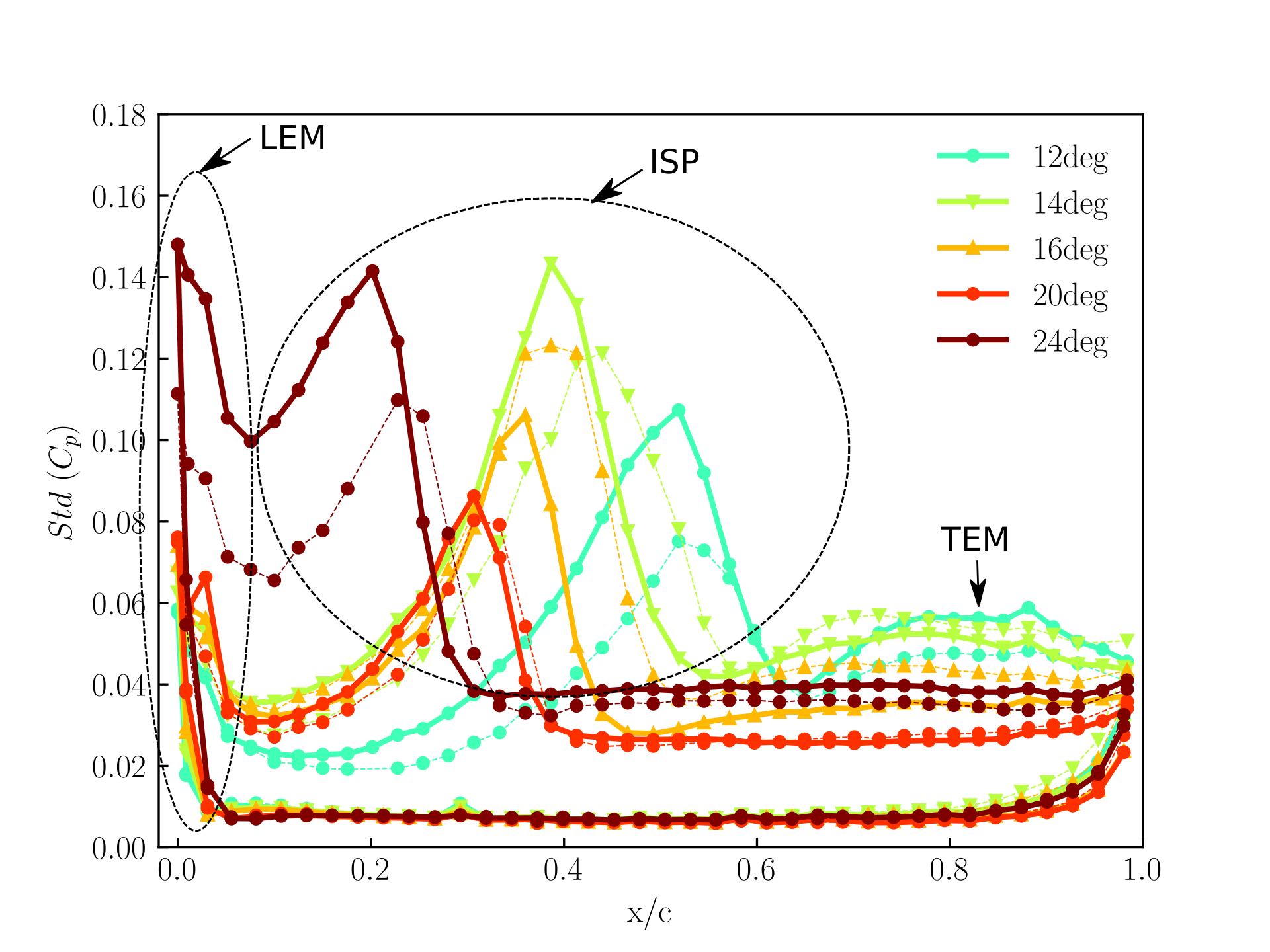}  
\end{tabular}
\caption{Standard deviation of the wall pressure coefficient, $Std(Cp)$, along the chord $x/c$, for both spanwise locations ($Y^{+}$ in dotted lines and $Y^{-}$ in solid lines), for different angles of attack: a) from \ang{6} to \ang{12}, b) from \ang{12} to \ang{24}. The markers indicate the pressure measurement locations. Note the difference of the scale for the y-axis between the low angles of attack (a) and high angles of attack (b) to highlight the apparitions of local maxima at \ang{10}. \berengere{ Local maxima are respectively denoted LEM: Leading Edge Maximum, ISP: Intermittent Separation Point, TEM: Trailing Edge Maximum}}
\label{fig:STD-CP}
\end{figure}

We now examine the spatial distribution of the wall pressure along the chord. 
Figures \ref{fig:CPmean} and \ref{fig:STD-CP} respectively represent the average and the standard deviation of the pressure coefficient for the different angles of attack.
The dotted lines in figures~\ref{fig:CPmean} and \ref{fig:STD-CP} represent the other row of pressures, $Y^+$. 
We note that the asymmetry observed in $C_N$ is also observable  for $C_P$ and $Std(C_P)$ quantities at the same angles of attack.

\subsubsection{Chord-wise wall pressure distribution}
\begin{figure}[htbp]
    \centering
    \includegraphics[width=0.55\textwidth]{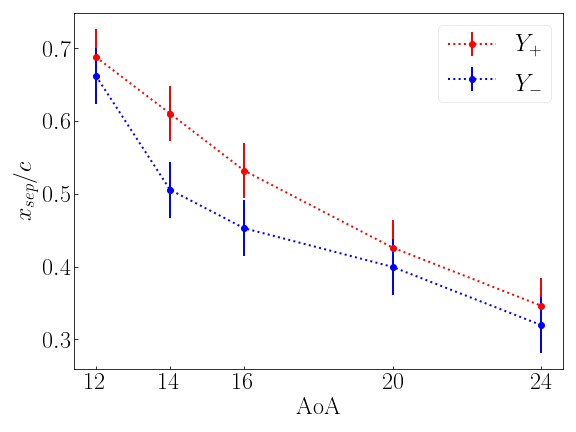}  
    \caption{Evolution of the separation point with the angle of attack for both spanwise locations. Error-bars represents the accuracy in the evaluation of the separation point due to the pressure tap resolution.}
    \label{fig:separationpt}
\end{figure}

The pressure distributions for all angles of attack have similar features: a peak of suction at the leading edge, a zone of high suction in the first half of the chord, then a decrease of the suction down to the trailing edge also called the recovery region.
The peak of suction at the leading edge increases with the increase of the angle of attack, reaching \num{-3} at an angle of attack of \ang{24}.
After the peak of pressure at the leading edge, the high suction zone loses intensity when the angle of attack increases, so that the recovery region starts closer to the leading edge.
A progressive flattening of the mean pressure coefficient curve is observed in the trailing edge region from \ang{12} to \ang{24}, which indicates that the flow is progressively separating according to the criteria of \citet{celik22, neunaber2022} (see section \ref{sec:separation_point}).
\berengere{Figure \ref{fig:separationpt} shows that the separation point moves progressively at a nearly constant rate towards the leading edge as the angle of attack increases, with a marked asymmetry between the spanwise locations at \ang{14} and \ang{16}. 
At other angles of attack, the observed differences fall within the uncertainty due to the spacing between sensors.} 


\subsubsection{Wall pressure fluctuations}
\begin{figure}[htbp]
\begin{center}

\includegraphics[width=0.7\textwidth]{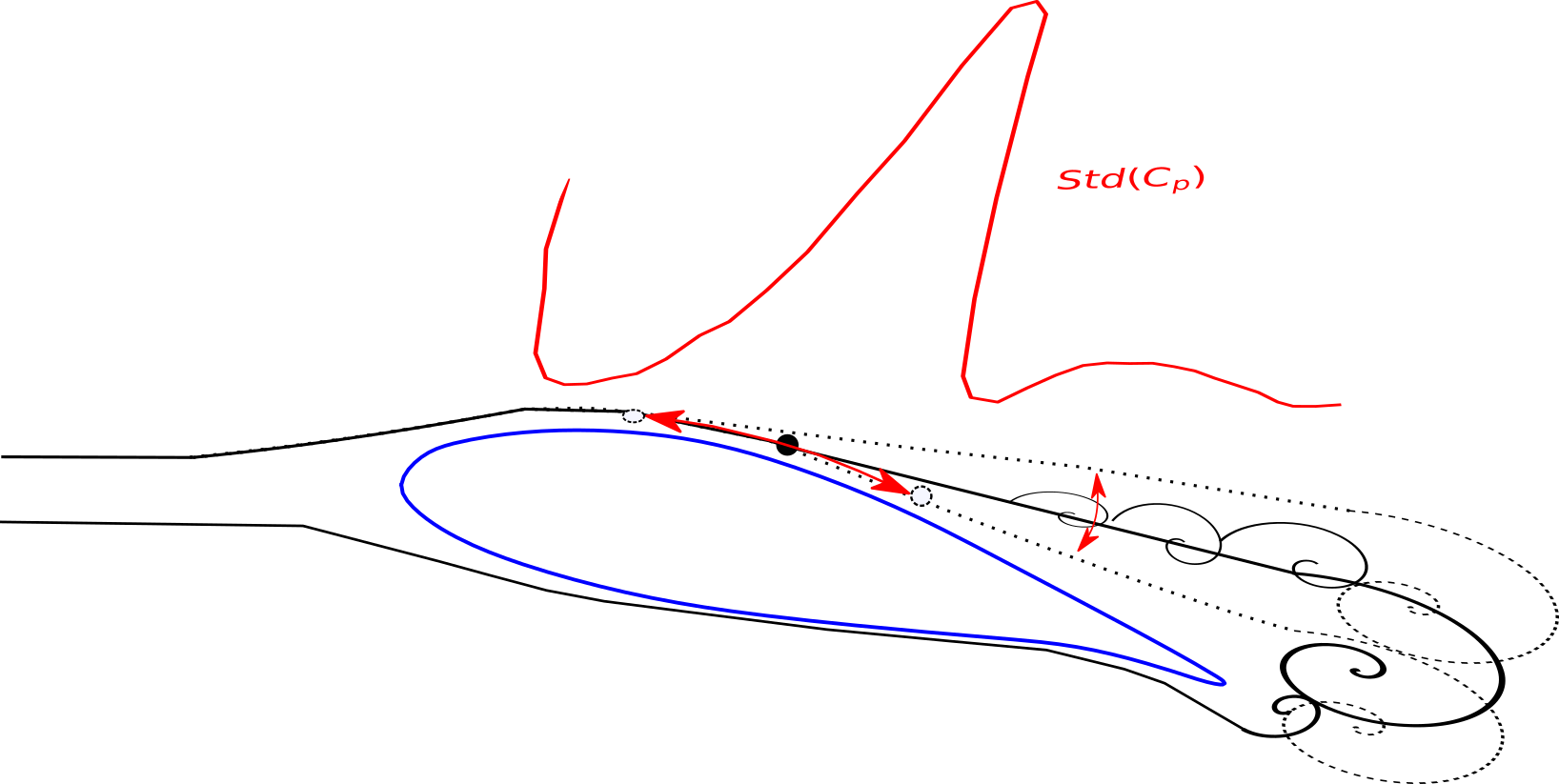}  
\end{center}
\caption{\berengere{Sketch of the flow physics corresponding to the spatial distribution of the pressure fluctuations. The red arrows along 
the blade surface show the intermittent flow separation associated with a large pressure standard deviation at the ISP, 
while the red arrows  above the blade represent the fluctuations of the shear layer associated with the TEM.}} \label{fig:scheme-bumps}
\end{figure}

\berengere{Figure \ref{fig:STD-CP} shows that the increase in the force fluctuations shown in figure~\ref{fig:CN} is characterized by the emergence of three local maxima: one close to the leading edge (denoted LEM for Leading Edge Maximum), a sharp one which we defined in section \ref{sec:separation_point} as the Intermittent Separation Point (ISP), and a smaller one corresponding to the separation region extending to the trailing edge, that is denoted TEM (Trailing Edge Maximum).  
The two last maxima (ISP and TEM in figure \ref{fig:STD-CP}a) are already noticeable at \ang{10}, which can be considered to represent a precursor state for flow separation over the airfoil.

\paragraph{Leading Edge Maximum, LEM}
The first maximum corresponds to the fluctuations of the suction peak at the leading edge and increases as the suction peak increases with the angle of attack.

\paragraph{Intermittent Separation Point, ISP}
The second maximum of fluctuations (ISP in figure \ref{fig:STD-CP}a) first appears at approximately 65\% of the chord at \ang{10}. 
This peak (ISP) is related to the intermittent displacement of the separation point as sketched in figure \ref{fig:scheme-bumps} by a red arrow around the ISP point on the blade surface. 
The instantaneous signal associated to this peak will be described further in section \ref{sec:pressure_jumps}.
At \ang{12}, when flow separation appears on the airfoil, the ISP moves at around 55\% of the chord and increases with an amplitude five times higher than at \ang{10}.
When the angle of attack is further increased until \ang{20}, the ISP decreases progressively in amplitude and moves towards the leading edge (see figure~\ref{fig:STD-CP}b).
At the highest angle of attack \ang{24}, the second maxima of pressure fluctuations (ISP) is located at 20\% of the chord and interacts with the LEM, with a significant increase of it.
High-intensity fluctuations are thus generated, which seems to signal the onset of a new regime corresponding to fully separated flow, and which will not be studied here.

\paragraph{Trailing Edge Maximum, TEM}
The third maximum in the trailing edge region (TEM at around 80\% of the chord) does not move but spreads out and progressively decreases in amplitude with the angle of attack until it totally disappears at \ang{20}.
The TEM is linked to the separated shear-layer whose distance from the airfoil oscillates, causing pressure variations as explained with figure \ref{fig:scheme-bumps}.
The centerline of the separated shear-layer is progressively moving away from the airfoil surface with the angle of attack which 
induces a decrease of the amplitude \citep{deparday_experimental_2022}.

}

\subsubsection{Pressure gradient in the intermittent separation region} 

\begin{figure}[htbp]
\begin{tabular}{cc}
\includegraphics[width=0.4\textwidth]{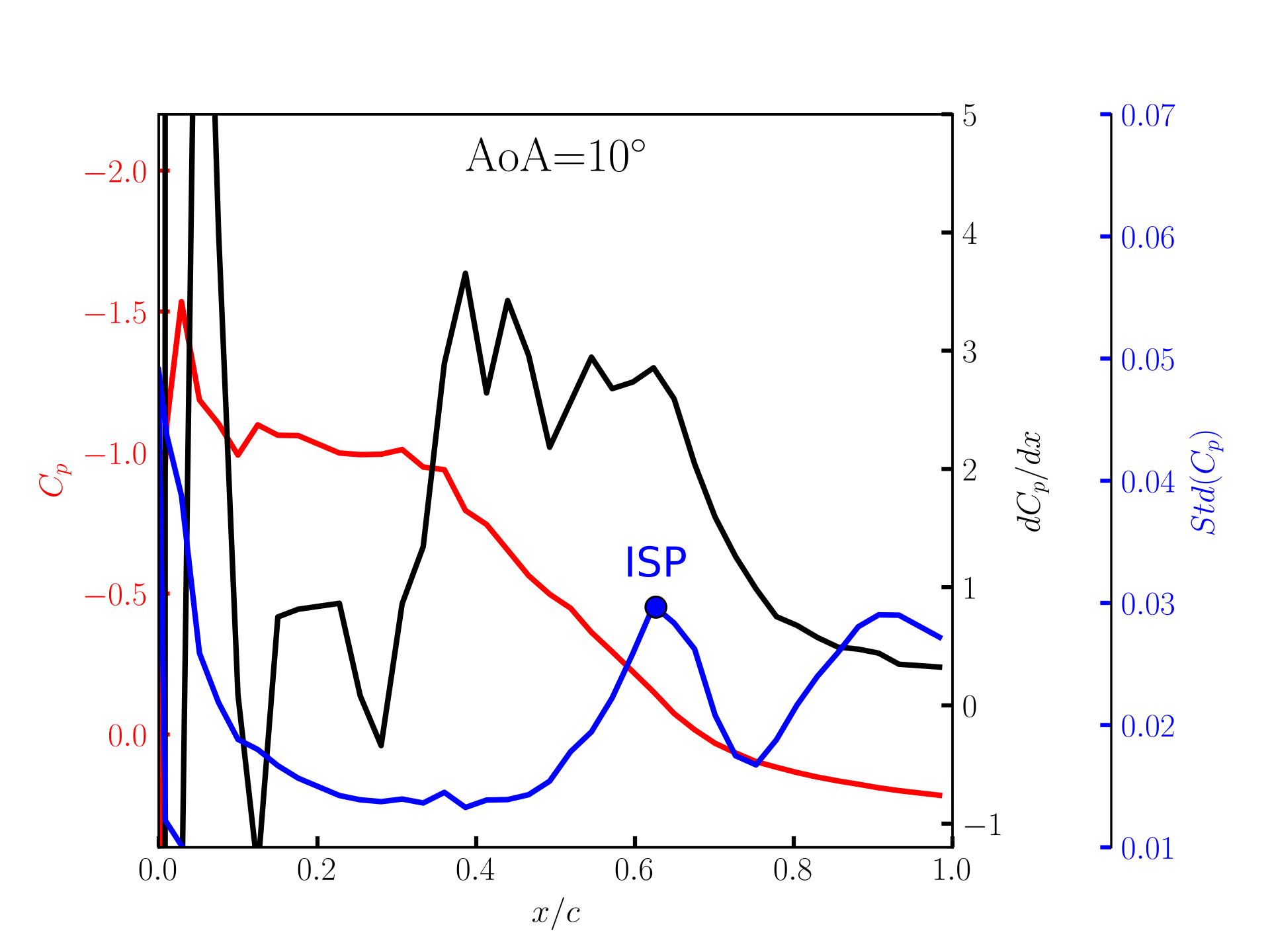}  &
\includegraphics[width=0.4\textwidth]{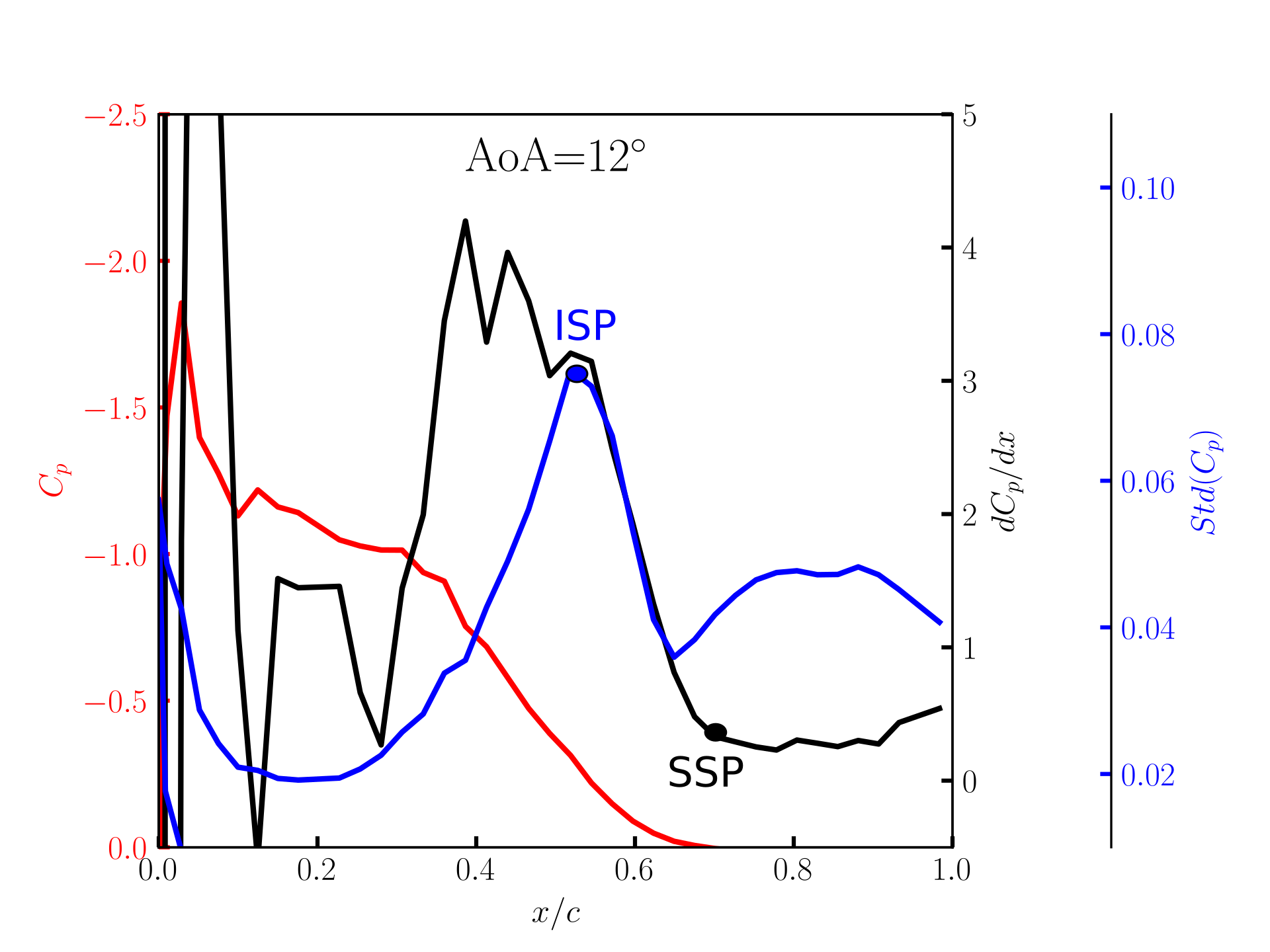}  \\
\includegraphics[width=0.4\textwidth]{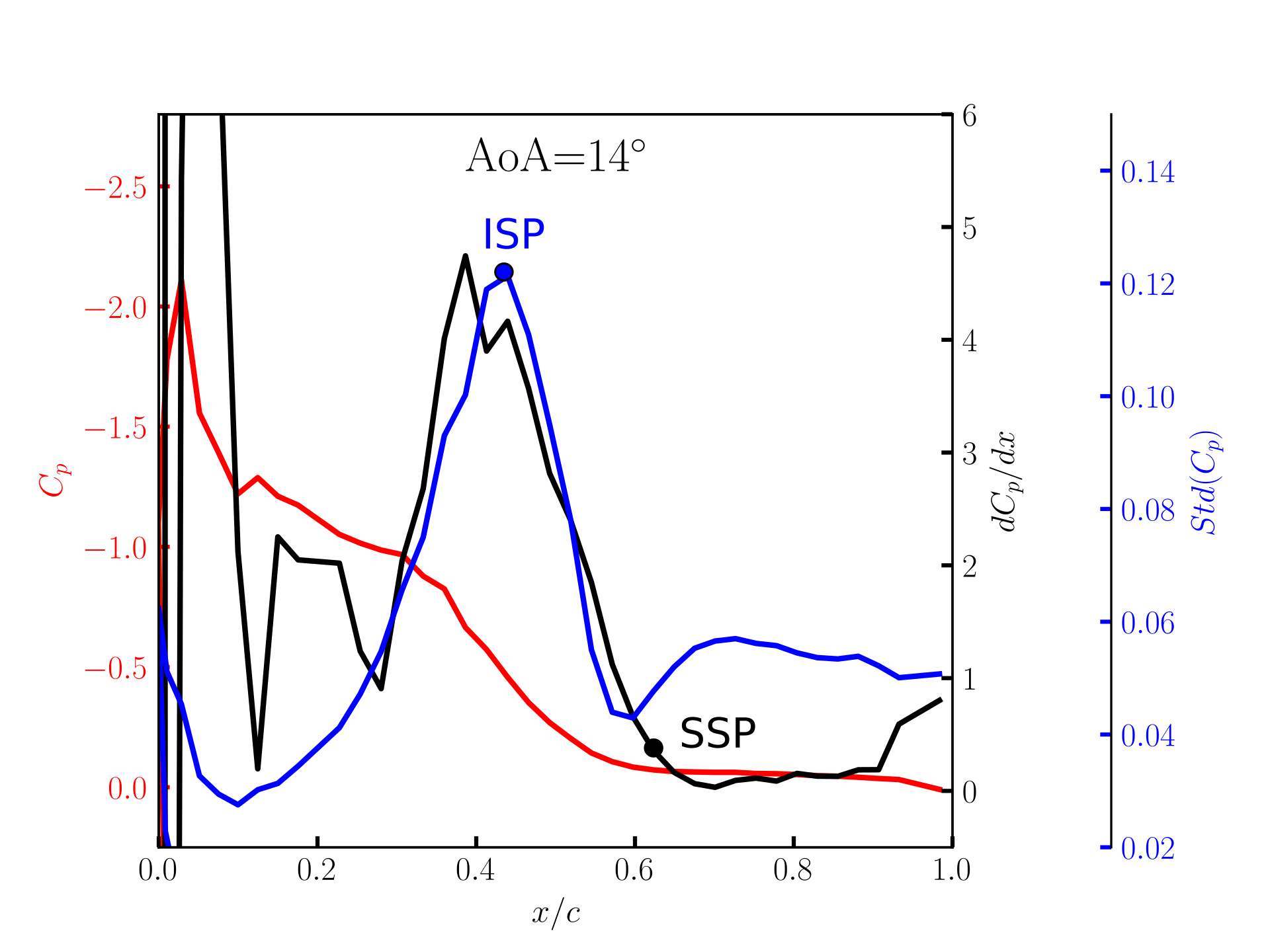}  &
\includegraphics[width=0.4\textwidth]{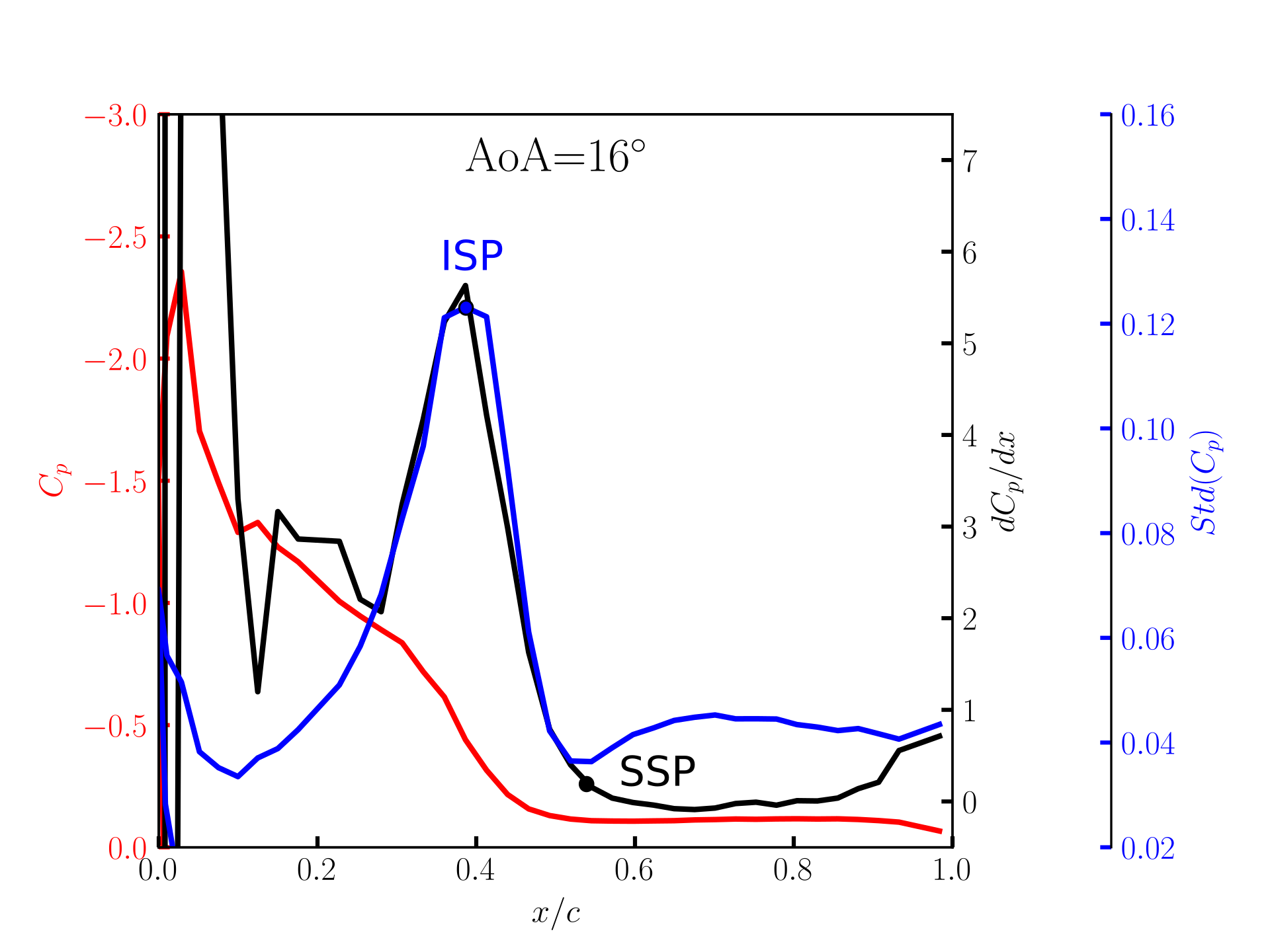}  \\
\includegraphics[width=0.4\textwidth]{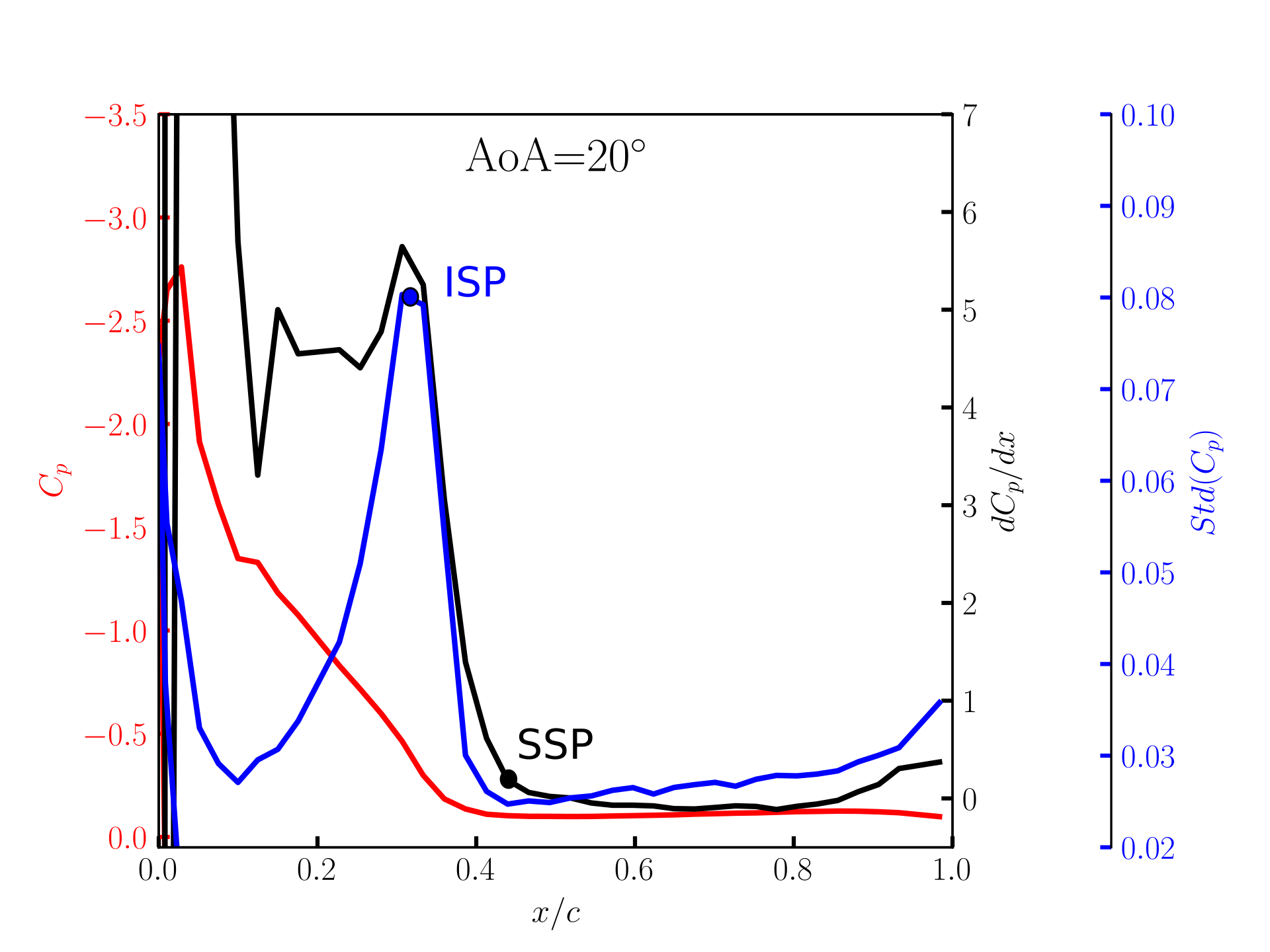}  &
\includegraphics[width=0.4\textwidth]{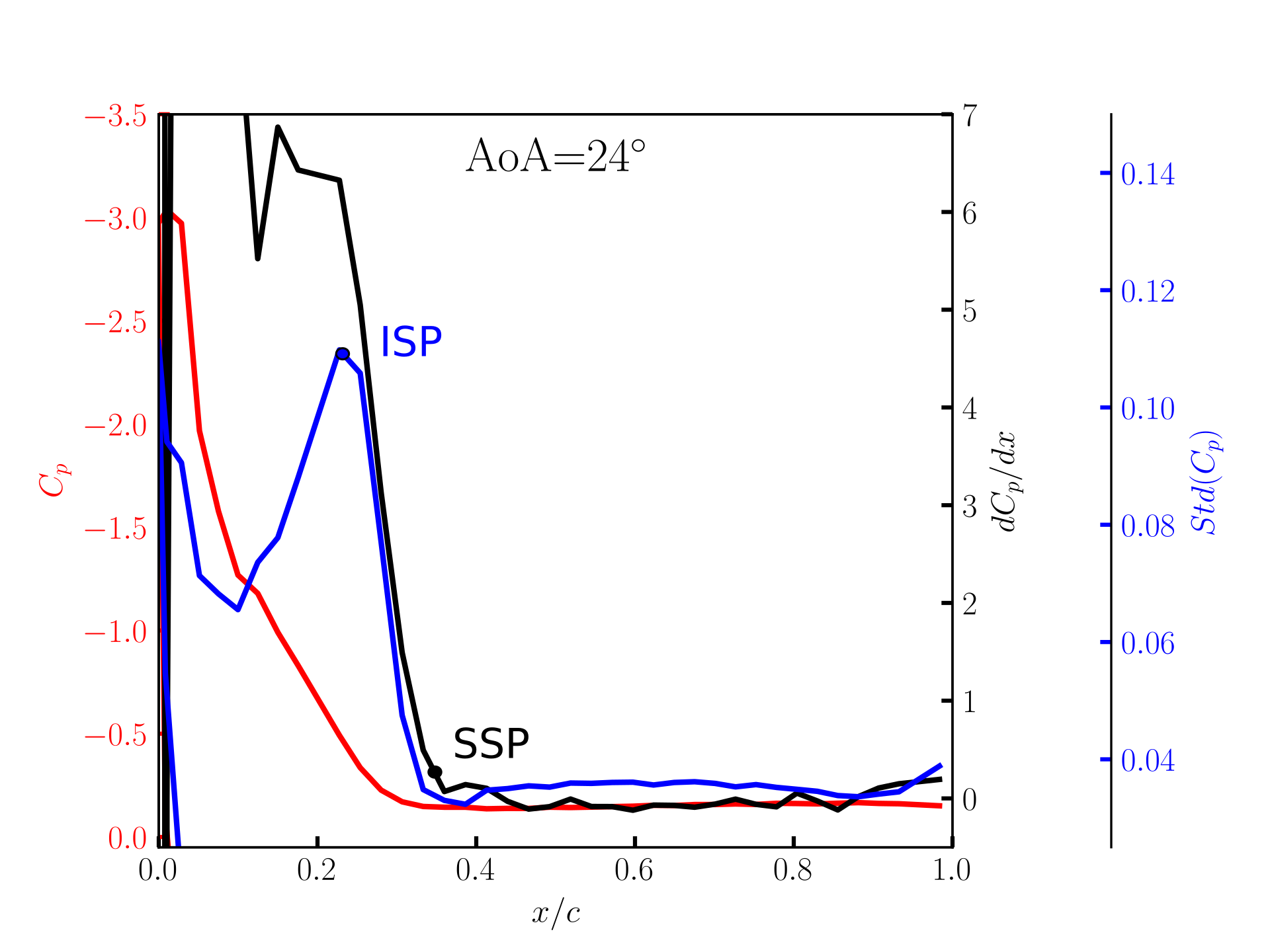} 
\end{tabular}
\caption{Pressure coefficients $C_p$ (in red), gradient of pressure $dC_p/dx$ (in black) and standard deviation of the pressures $Std(Cp)$ (in blue) versus the chord-wise direction, $x/c$ at the $Y^-$ row for different angles of attack (from \ang{10} to \ang{24}). \berengere{ The intermittent separation point (ISP) and the steady separation point (SSP) locations are indicated in the figures.}}
\label{fig:dCPmean2}
\end{figure}

\berengere{  In the literature, the available quantity is generally limited to the average pressure around the chord at 
one blade span location. It is thus interesting to underline the link between the mean pressure and the fluctuating pressure that is 
related to the instability described in the present paper.} 
Figure~\ref{fig:dCPmean2}
 displays on the same plot the pressure gradient (in black), the pressure standard deviation (in blue) and the mean pressure coefficient (in red) for the blade suction side. 
We will focus on the intermittent separation region away from the leading edge.
At \ang{10},  the mean pressure gradient takes high values in a large region at mid chord spans from $x/c = 0.35$ to $x/c = 0.7$.
The width of the region  progressively shrinks with the increase of the angle of attack until \ang{14} where 
a single peak is defined. 
As reported earlier, at \ang{10}, the wall pressure standard deviation peak (in blue) displays two maxima: 
one in the central region corresponding to the large adverse pressure gradient  and one close to the trailing edge. 
As the angle of attack increases, the trailing edge maximum decreases and disappears for angles higher than \ang{16}, while
that in the pressure gradient region becomes important for \ang{14} and \ang{16} i.e. where load fluctuations are maximal.
A clear coincidence between the mean pressure gradient and standard deviation peaks can be observed for these angles.
The strong connection between the two regions is still present at higher angles, as both peaks move toward and 
progressively merge with  those at the leading edge. 
coincide 
As detailed in section \ref{sec:separation_point}, the steady separation point (SSP) is determined as the location where 
the pressure gradient falls below a certain threshold, while the intermittent separation point (ISP) is defined at the peak of 
pressure fluctuations (see figure \ref{fig:separationpt}). These points are reported in figure \ref{fig:dCPmean2}. It is noticeable that 
for all angle of attacks over \ang{12}, the ISP is located directly upstream of the SSP similarly as reported by \cite{simpson87}.\\

In summary, it is shown that pressure fluctuations are significant in  three regions: i) the leading edge, ii) the strong adverse pressure gradient region corresponding to the occurrence of intermittent separation and iii) the area where the flow is fully detached at the trailing edge.
The pressure fluctuations are the highest around the maximal lift conditions, with a global fluctuation maximum in the intermittent separation region (ISP) that coincides with the local pressure gradient maximum.

\subsection{Coherence of the pressure fluctuations}

\begin{figure}[htbp]
\includegraphics[width=1\textwidth]{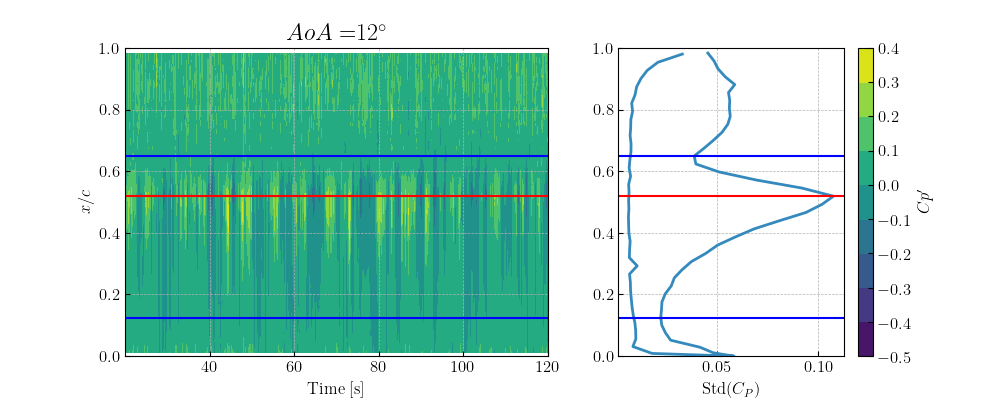}  
\caption{Left: Spatio-temporal evolution of the wall pressure fluctuations at the spanwise location $Y^{+}$ (only suction side) and for $AOA=12^\circ$. Right:\caro{Wall pressure standard deviation} distribution for $AOA=12^\circ$. For both plots, the red line represents the maximum location of the standard deviation peak in the recovery region and the blue lines correspond to the two local fluctuation minima  delimiting the region of strong fluctuations.}
\label{fig:instCPYP}
\end{figure}

\begin{figure}[htpb]
\begin{tabular}{cccc}
&\includegraphics[width=0.3\textwidth]{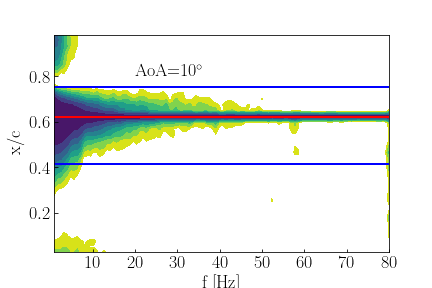}  &
\includegraphics[width=0.3\textwidth]{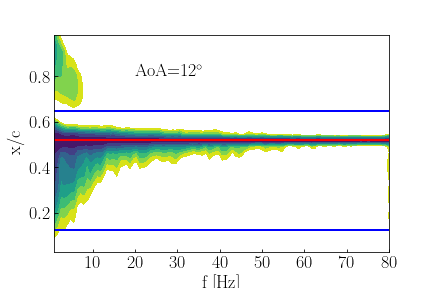}  &
\includegraphics[width=0.3\textwidth]{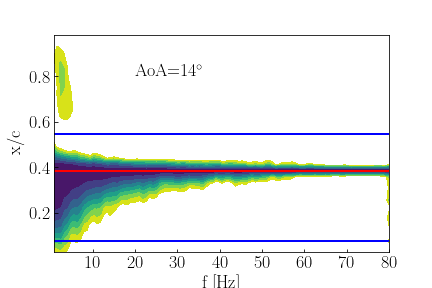}  \\
\includegraphics[width=0.1\textwidth]{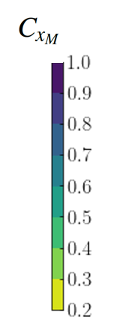}  &
\includegraphics[width=0.3\textwidth]{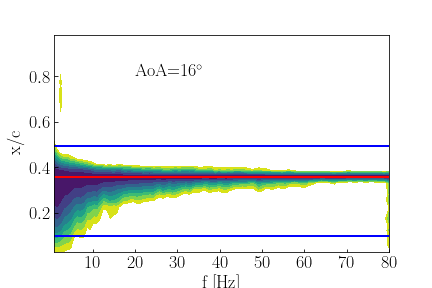}  &
\includegraphics[width=0.3\textwidth]{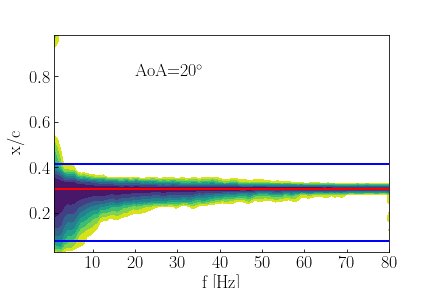}  &
\includegraphics[width=0.3\textwidth]{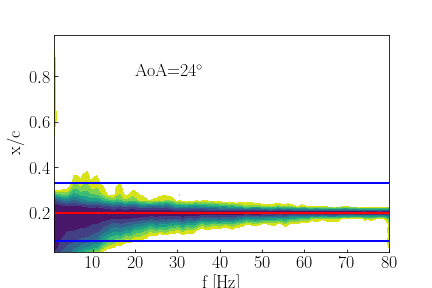}  \\
 
\end{tabular}
\caption{Coherence $C_{x_M}(x,f)$ between the  pressure coefficient at location on the suction side and the intermittent separation point  $x^{ISP}$ (indicated with a red line) on the chord $Y=Y^+$. The frequency is indicated in Hertz. The blue lines correspond to local minima of the pressure fluctuations. 
  }
\label{fig:coherenceYp}
\end{figure}

\berengere{
The spatio-temporal evolution of the  pressure fluctuations is represented at $AoA=12^\circ$ in Figure~\ref{fig:instCPYP} with the corresponding standard deviation.
The local standard deviation minima (blue lines on the figure) are unambiguous criteria to delimit the region of strong pressure fluctuations that will be identified from now on as  the intermittent separation region.
The spatio-temporal map shows that at most instants, a strong spatial correlation is apparent in the intermittent separation region (and to some lesser extent around the trailing edge). 
Some correlation in time is also noticeable in the intermittent separation region, in particular around the ISP.

To characterize the organization of the fluctuations, we computed the coherence function of the pressure coefficient between the intermittent separation point (ISP) and other chordwise location $x/c$. 
The coherence function is defined as  
\begin{equation}
C_{x^{ISP}}(x,f)= \frac{S_{xx^{ISP}}(f)}{S_{xx}^{1/2}(f)S_{x^{ISP}x^{ISP}}^{1/2}(f)}, 
\end{equation}
where $x$ is a location on the chord, $x^{ISP}$ is the intermittent separation point (aka maximum fluctuation  location), $S_{xx^{ISP}}$ represents the cross-spectral density  of the pressure signal at locations $x$ and $x^{ISP}$ and $S_{xx}$ represents the power spectral density at location $x$.
High values of the coherence function represented in figure \ref{fig:coherenceYp} can help identify spatial regions over which the pressure signals are well correlated. 
For clarity, the positions of the local maximum and local minima (respectively blue and red lines) are also reported in figure \ref{fig:coherenceYp}.
At an angle of \ang{10}, the main coherent region can be observed both upstream and downstream of the ISP. Its extent broadly coincides with the definition of the intermittent separation region as delimited by the blue lines.  
Some coherence is also present at the trailing edge, but its intensity gradually decreases and essentially disappears for $AoA \ge \ang{16}$. 
After separation, the main coherent region is relatively more important upstream of the maximum location. 
Its size tends to decrease as the angle of attack increases.
We note that no clearly identified frequency appears in the coherence plots, which seem dominated by a mixture of low frequencies ($\le 20 Hz$).}


\section{Local pressure fluctuations}
\label{sec:local}


As seen in Figure~\ref{fig:CN} and figure~\ref{asym}, it has been confirmed that the pressure variations and the separation points differ for the two rows of pressure $Y^{+}$ and $Y^{-}$ at angles of attack between \ang{12} and \ang{20}.
In \citet{neunaber2022}, it was found that the highly unstable displacements of the separation point along each chord were anti-correlated, and could be associated with a bistability phenomenon.
The bistability was characterized in two ways: strong pressure {\bf temporal jumps} on each chord, associated with {\bf spatial switches} in intensity from one chord to another. 
The jumps and switches are investigated below using the instantaneous wall pressure signals at the ISP 
, which evolves with the angle of attack. 

\subsection{Jumps}
\label{sec:pressure_jumps}

\begin{figure}[htbp]
\includegraphics[width=0.95\textwidth]{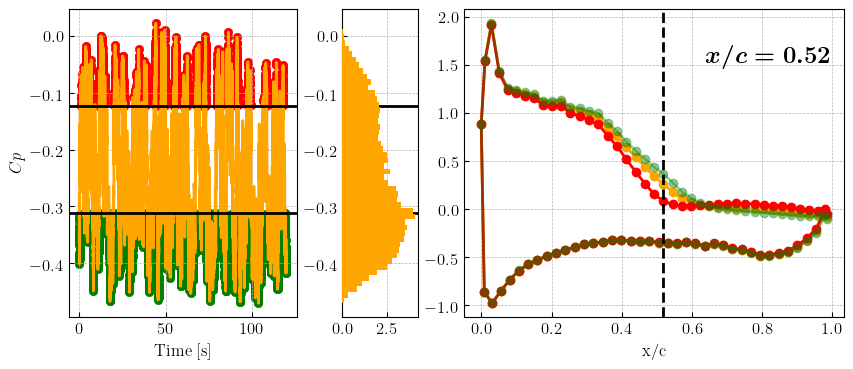} 
\caption{Left: $AoA=12^\circ$, Instantaneous pressure signal ($Y^{+}$) at $x/c=0.52$ at the intermittent separation point; Center:  Histogram of the pressure signal. Right: Averaged pressure coefficient distribution based on full and conditional averages:  high values (red), low values (green) and all values (orange).}
\label{fig:jump12deg}
\end{figure}

In this section, we attempt to characterize the pressure jumps observed in the fluctuations, which is not straightforward given the temporal complexity of the signal (see figure \ref{fig:coherenceYp}). 
To do so, we consider the instantaneous pressure signal at the ISP on a single chord (we choose $Y^{+}$)  at the angle of attack \ang{12}, shown in 
figure \ref{fig:jump12deg}-left. 
The associated histogram (figure \ref{fig:jump12deg}-center) shows a well-defined local maximum in the low-value range and a less clear one in the high-value range. 
These maxima (corresponding to the thick horizontal black lines in the figure) can be used 
as thresholds to delimit low-value  and high-value  regions (indicated respectively in green and red on the figure). 
A jump can then be defined as an excursion of the signal between the low-value and the high-value region. 
In figure \ref{fig:jump12deg}-right), conditional averages of the pressure coefficient based on these two regions are compared with the average pressure coefficient distribution (in orange). At the ISP location, materialized by the dashed vertical line, low-value regions correspond to attached flow and high-value regions to separated flow.

\begin{figure}[htbp]
\begin{tabular}{cc}
a)\includegraphics[width=0.5\textwidth]{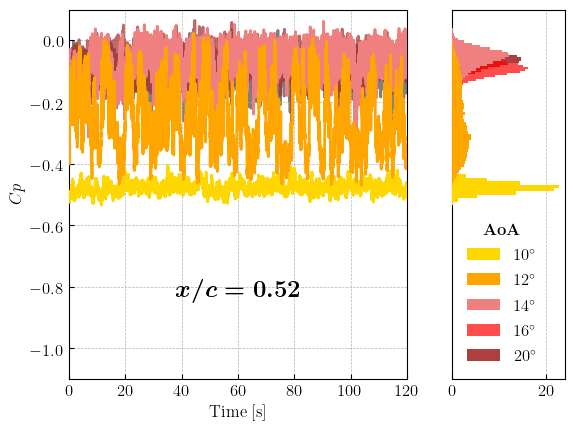}  &
b)\includegraphics[width=0.5\textwidth]{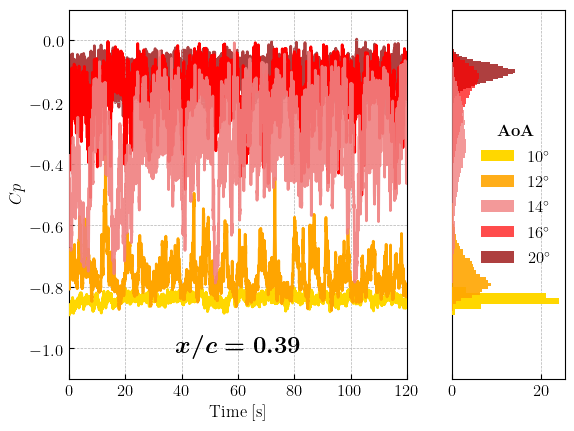} \\ 
c)\includegraphics[width=0.5\textwidth]{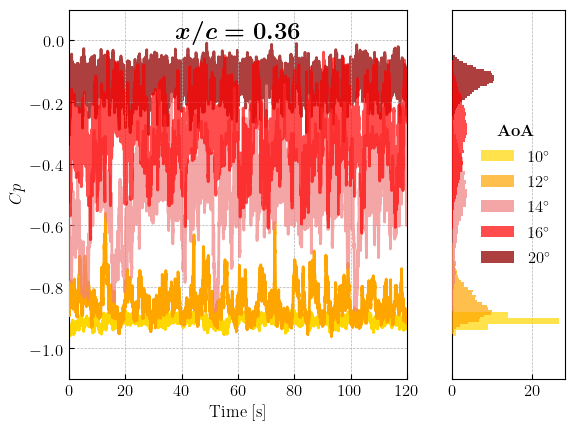} & 
d)\includegraphics[width=0.5\textwidth]{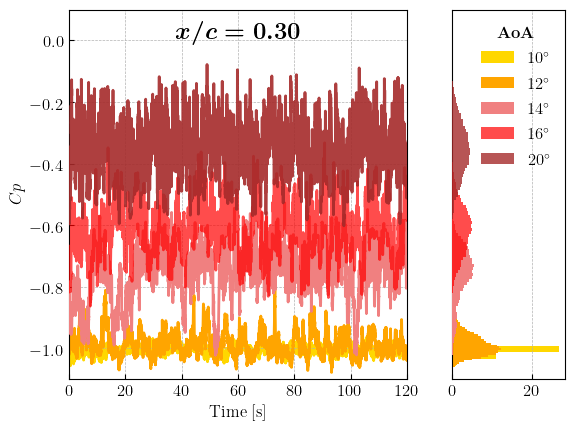} 
\end{tabular}
\caption{Instantaneous pressure signals ($Y^{+}$) for different angle of attacks at 4 chordwise locations, $x/c=$0.52, 0.39,  0.36 and 0.30, corresponding to the intermittent separation points at respectively \ang{12}, \ang{14}, \ang{16} and \ang{20}. The associated histograms are plotted on the right-hand side of each figure.}
\label{fig:CP_jumps}
\end{figure}

To highlight this evolution, the instantaneous pressure signal is observed at a fixed point for different angles of attack (including the intermittent separation point).
For instance, in figure \ref{fig:CP_jumps}.a), the chord location corresponds to the intermittent separation point at \ang{12}, where the pressure variations are the highest compared to those at other angles of attack (\ang{10}, \ang{14}, \ang{16} and \ang{20}).
The pressure signal at \ang{12} jumps from one mean level measured for the angle of attack of \ang{10}, where the flow is attached at this chord location, to another mean level measured for the higher angles of attack, where the flow is detached. 
Instead of oscillating around a mean value located between the mean levels of \ang{10} and \ang{14}, with small amplitudes similar to those measured at the other angles of attack, the pressure signal at \ang{12} has large variations, making large excursions into these two levels. 
The histogram plotted on the right side shows an elongated distribution with tails reaching both attached and separated pressure levels.
The distribution has an almost bimodal shape with a negative skewness, which indicates that for this angle of attack, even if the flow is switching between attached and separated flow states, the preferred state is the attached flow.

For the other chord locations, from figure \ref{fig:CP_jumps}.b) to figure \ref{fig:CP_jumps}.d), the pressure difference increases significantly compared to figure \ref{fig:CP_jumps}.a) (see the difference between the brown curve and the yellow curve) and keeps increasing when the intermittent separation region moves toward the leading edge. 
In figure \ref{fig:CP_jumps}.b), corresponding to the intermittent separation point for an angle of attack of \ang{14}, the wall pressure signal evolves from a negative (\ang{12}) to a positive (\ang{14}) skewness, 
so from a flow mostly attached to a flow mostly separated with no intermediate states within the \ang{2} 
step of the angle of attack
\berengere{(note that on the other chord, the flow remains mostly attached at \ang{14})}.
For figure \ref{fig:CP_jumps}.c), corresponding to the intermittent separation point for an angle of attack of \ang{16}, the pressure signal at \ang{14} displays a symmetric distribution representing an intermediate state between the mostly attached and mostly detached state.
The pressure signal at \ang{16} is flatter and reach both this intermediate state and the mostly detached state.
For figure \ref{fig:CP_jumps}.d), where the intermittent separation point is defined for an angle of attack of \ang{20}, no narrow and well defined detached state is easily defined, and the pressure signals for \ang{14} and \ang{16} show symmetric distributions, representing intermediate states.
As a result, bistability  cannot be easily identified anymore.

\subsection{Switches}

\begin{figure}[htbp]
\begin{tabular}{cc}
a)\includegraphics[width=0.5\textwidth]{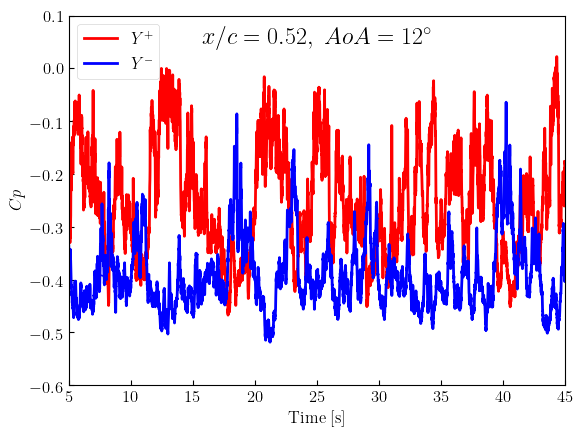}  &
b)\includegraphics[width=0.5\textwidth]{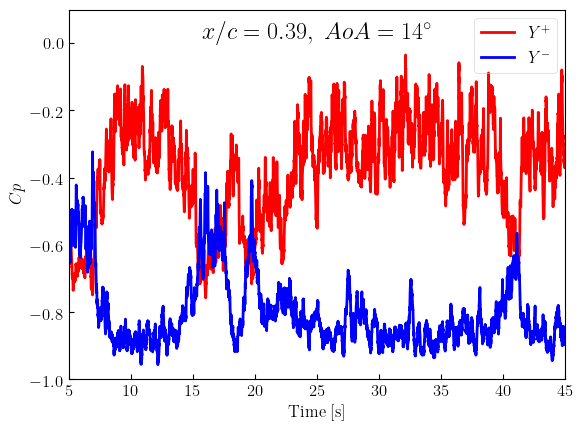}  \\
c)\includegraphics[width=0.5\textwidth]{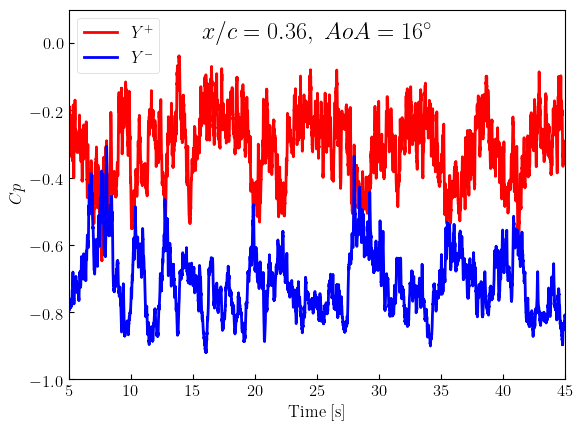}  &
d)\includegraphics[width=0.5\textwidth]{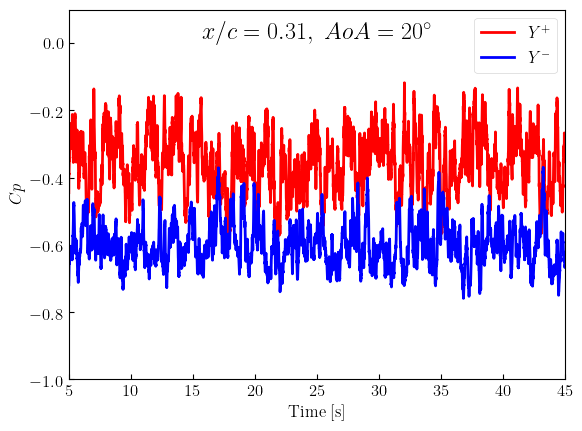}  
\end{tabular}
\caption{Spatial pressure switch between $Y^{+}$ in red to $Y^{-}$ in blue for different chord-wise locations corresponding to the intermittent separation point of the considered angle of attack}
\label{fig:CP_switch}
\end{figure}

In this section, pressure signals at both locations $Y^{+}$ and $Y^{-}$ are examined.
A  switch  (see figure \ref{fig:CP_switch}) is defined as a moment where a large pressure jump is observed on one side, $Y^{+}$ (or $Y^{-}$), while coinciding with a pressure jump of the opposite sign at the other spanwise location, $Y^{-}$ (or $Y^{+}$). 
The switches between $Y^{+}$ and $Y^{-}$ can be seen in the pressure signals measured at the intermittent separation point locations, which are shown in figure \ref{fig:CP_switch} for angles of attack ranging from \ang{12} to \ang{20}.
The switches are clearly apparent for the angles of attack \ang{12} (figure \ref{fig:CP_switch}.a)) and \ang{14}(figure \ref{fig:CP_switch}.b)), and are correlated with sharp jumps from an attached to partially detached flow shown in figure~\ref{fig:CP_jumps}.
The  switches are not as well identified at \ang{20} (figure \ref{fig:CP_switch}.d)), which is consistent with the less clear jumps and the appearance of intermediate states described in the previous section.

\begin{figure}[htpb]
\centerline{ \includegraphics[trim=0cm 0 1cm 0, clip,height=10cm]{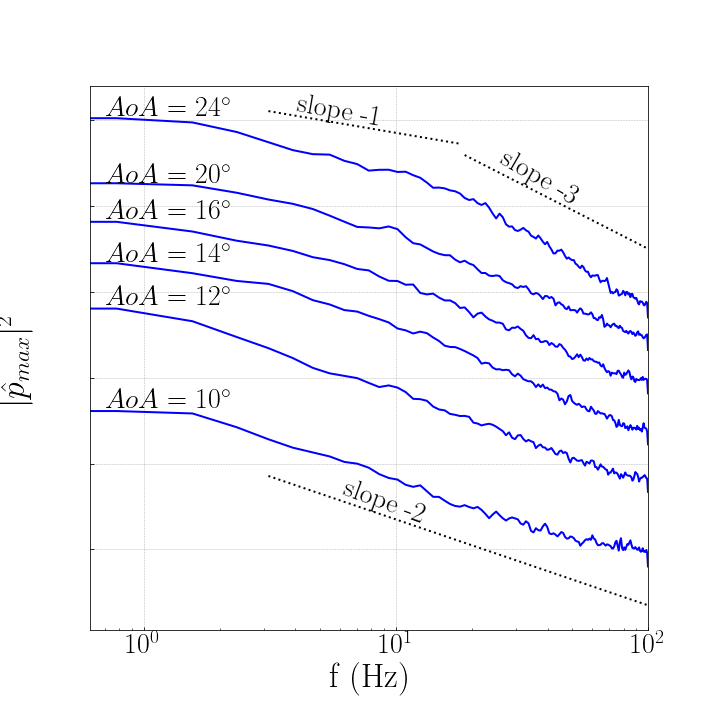} }
\caption{Spectra of the pressure signal at the intermittent separation point for different angles of attack (the spectra have been rescaled for easier comparison).} 
\label{spectrummax}
\end{figure}
 
\berengere{Figure \ref{spectrummax} shows the temporal spectrum of the pressure signal at the intermittent separation point for different angles of attack after the onset of separation. 
No single time scale could be directly identified in the pressure signal to characterize the jumps and switches of the bistability phenomenon.
We can see that it is consistent with the scenario of a gradual process involving a wide spectrum of time scales:
at high frequencies, one observes an evolution from a $f^{-2}$ behavior at \ang{12} to a $f^{-3}$ region between \ang{16} and \ang{20}. 
A sharper drop-off is observed at \ang{24}.}
We note that \citet{kiya82} found a scaling between -2 and -3 in the recovery region of the separated flow over a blunt flat plate, while a -3 spectrum for high frequencies was identified in several turbulent adverse pressure gradient flat-plate boundary layers \citep{namoin98} (see also \citet{simpson87} for a review).

\begin{figure}
\includegraphics[trim=0cm 0 0cm 0, clip,height=6cm]{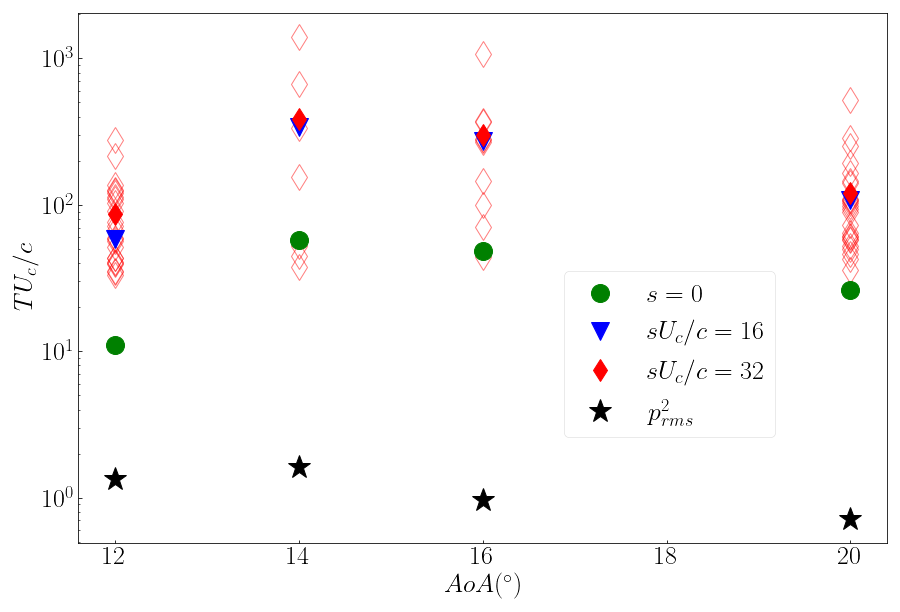}
\caption{Evolution of the bi-stability time scale with the angle of attack for different values
of the time threshold $s$.
The open symbols correspond to the different intervals measured for $sU_c/c=32$. The variance of the pressure fluctuations $p^2_{rms}$ is also represented on the plot for comparison as star markers.}
\label{timesc}
\end{figure}

Despite the absence of a well-identified cyclic process, a characteristic time scale can still be extracted from the average time interval between such jumps.
To do so, we  considered the signals at the intermittent separation point on each chord and determined the times when one became higher than the other (similar results were obtained when the normal force on the chord was used instead of the local pressure signal).
We defined the characteristic time scale $T$ to be the average duration between these events.
Events lasting less than a time threshold $s$ were excluded, to exclude possible transients.
Three different thresholds are represented in figure \ref{timesc}:  
$s=0$ (all events were included),  $s=$\SI{0.5}{\second} and $s=$\SI{1}{\second}.   
No significant changes were observed for the last two cases, which provides some confidence in the robustness in the results.
\berengere{Sample time intervals are also shown in figure \ref{timesc} as open symbols for a threshold of $s=4$, and it can be seen that the variations of the distribution with the angle of attack are well captured by  the characteristic time scale.  
In all cases where  bi-stability is present, the time scale seems to evolve roughly like the variance of the pressure fluctuations. }
%
\berengere{ This meaans that large pressure variations corresponding to the difference between  attached and separated flow states 
are less frequently observed than small pressure variations corresponding  to the existence of intermediate states.}  

\section{POD analysis}
\label{sec:pod}

\berengere{ To complement the local description of the previous section, we provide a low-order characterization of the global dynamics based on Proper Orthogonal Decomposition (POD). 
For each angle of attack, 
the decomposition was applied independently to each chordwise row of pressure taps - limited to the suction side. 
Only the fluctuating part of the signal was considered.}

\subsection{Eigenvalues and eigenmodes}

\begin{figure}
\centerline{ \includegraphics[trim=0cm 0 1cm 0, clip,height=10cm]{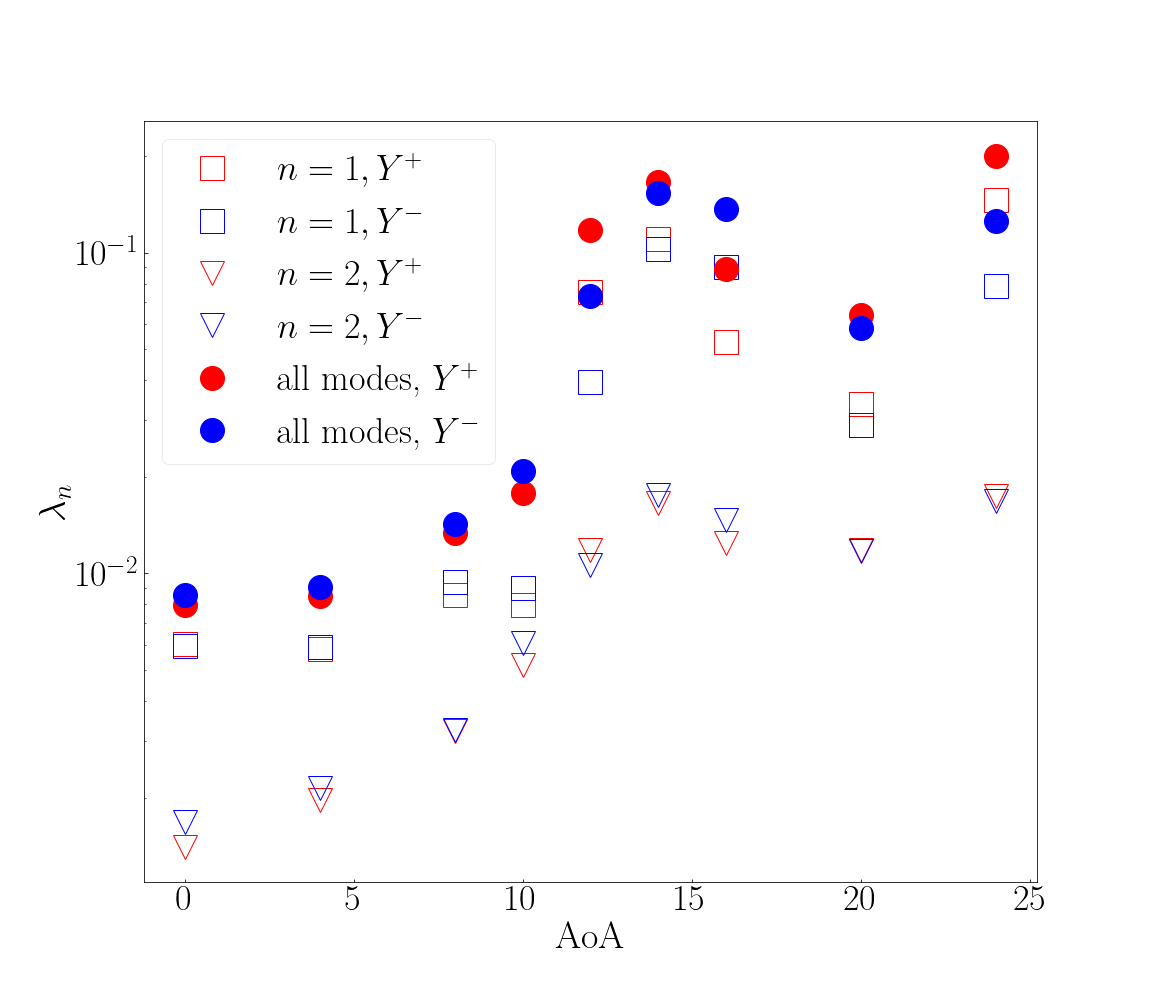} }
\caption{Evolution with the angle of attack of the first two POD eigenvalues $\lambda_1$ (squares) and $\lambda_2$ (triangles), referring to the energy of the two first modes and of the total variance (solid circles), for each row of pressure $Y^{+}$ (in red) and $Y^{-}$ (in blue).} 
\label{lambdan12}
\end{figure}

Figure \ref{lambdan12} shows the first two eigenvalues $\lambda_1$ (squares) and $\lambda_2$ (triangles) for both spanwise locations ($Y^+$ in red and $Y^-$ in blue) as function of the angle of attack.
The first two eigenvalues represent the two most energetic contributions to the total variance (full circles) (i.e energy of the fluctuations).
The total variance globally increases with the increasing angle of attack, but with a drop at \ang{20}.
Before the flow separation, the two modes capture respectively 70\% and 20\% of the variance.
After the start of the flow separation at an angle of attack of \ang{12}, over 70\% of the total variance is still captured by the first two modes, which suggests that the pressure dynamics could be successfully captured with a low-order representation.
Generally speaking, eigenvalues tend to increase following the trend of the total variance, with one exception: 
just before the start of the flow separation, at \ang{10}, although the total variance increases sharply, the energy of the first mode actually decreases and becomes close to that of the second mode, which points out to a strong reorganization of the fluctuations.
After flow separation, the energy of the first mode increases sharply again and a disymmetry between the two chords, i.e. $Y^+$ and $Y^{-}$, is observed.
The first mode at $Y^+$ appears to have more energy at \ang{12}, while $Y^-$ has more energy at \ang{16}, which is consistent with what is observed in figure~\ref{fig:STD-CP}.
Symmetry appears to be restored at \ang{14}, when the fluctuation level is maximum, as well as at \ang{20}. 
At \ang{24}, the symmetry is broken again. 
This evolution confirms that strong, three-dimensional reorganizations of the flow take place as the angle of attack varies.

\begin{figure}
 \renewcommand{\arraystretch}{0.8} 
\begin{tabular}{ccc}
   \small{$AoA=0 ^\circ$}  & \small{$AoA=4 ^\circ$ } & \small{$AoA=8 ^\circ$ } \\
 \includegraphics[trim=2.5cm 0 3.cm 0, clip, height=4.5cm]{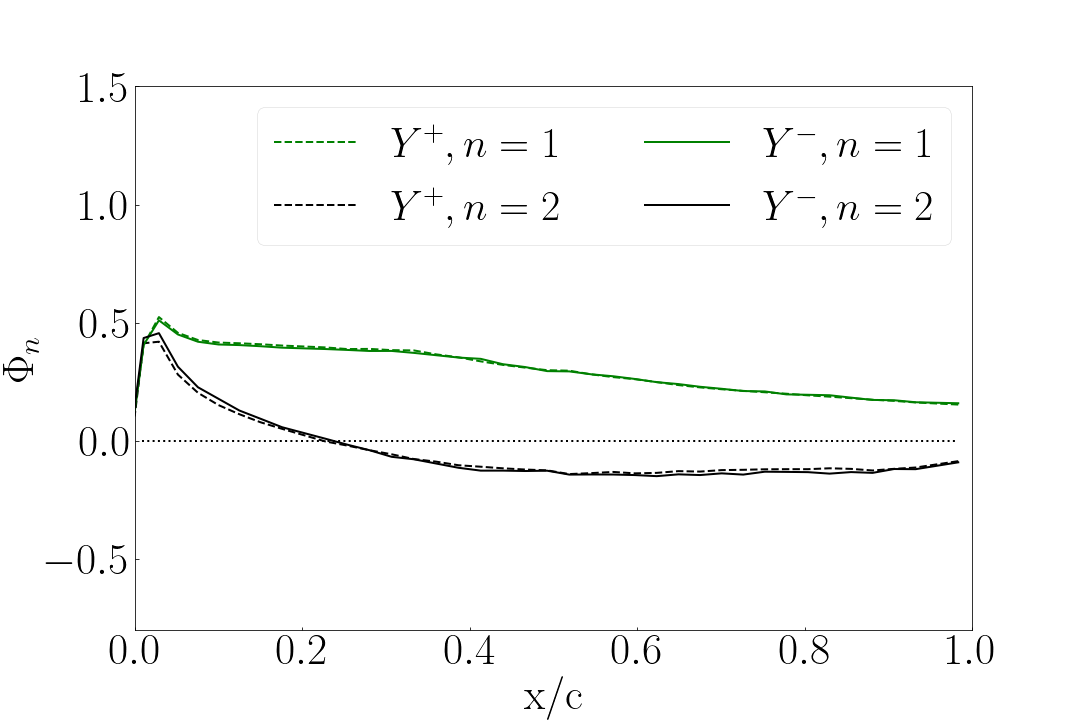} & \includegraphics[trim=2.5cm 0 3.cm 0, clip,height=4.5cm]{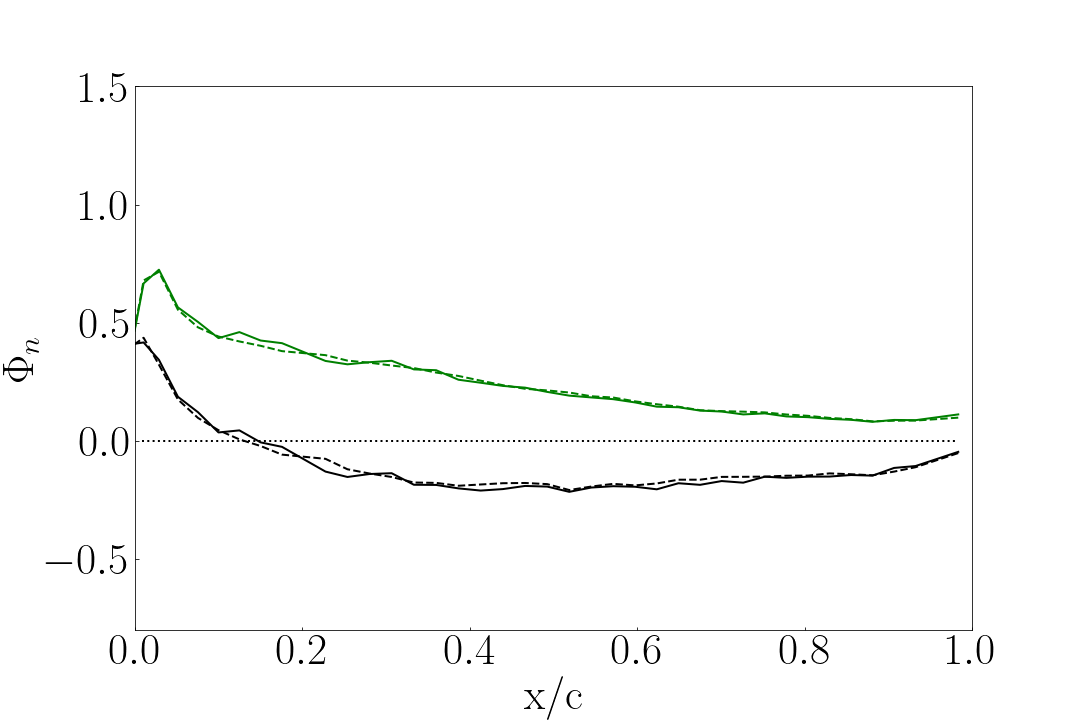} & \includegraphics[trim=2.5cm 0 3.cm 0, clip, height=4.5cm]{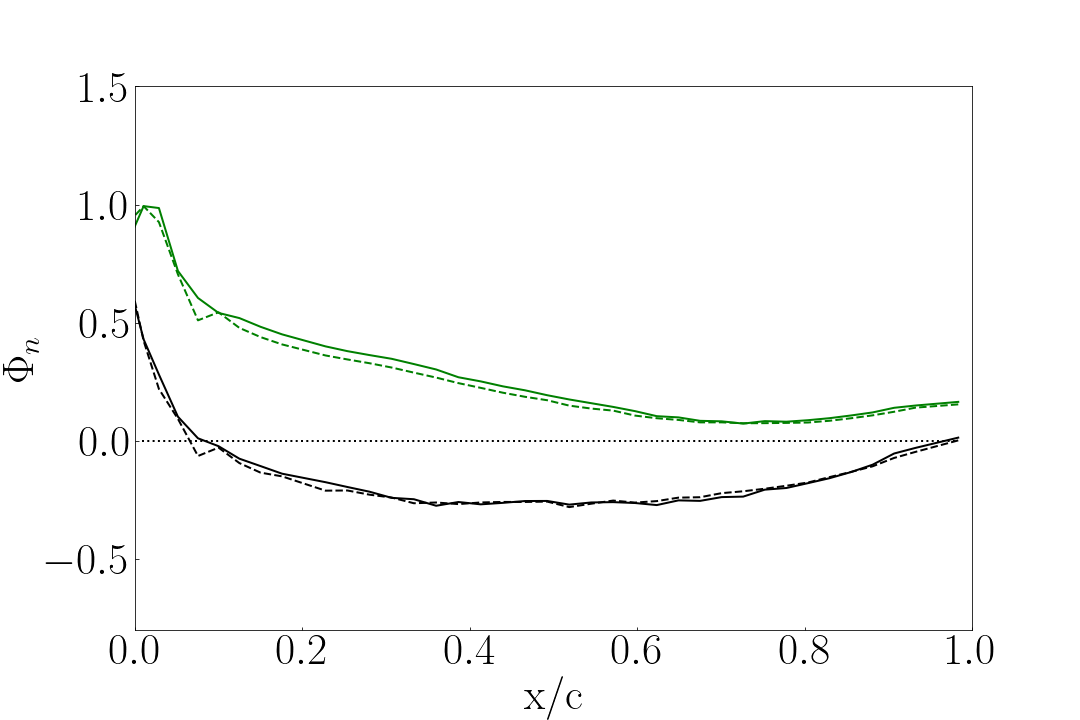} \\ 
   \small{$AoA=10 ^\circ$}  & \small{$AoA=12 ^\circ$ } & \small{$AoA=14 ^\circ$ } \\
 \includegraphics[ trim=2.5cm 0 3.cm 0, clip,height=4.5cm]{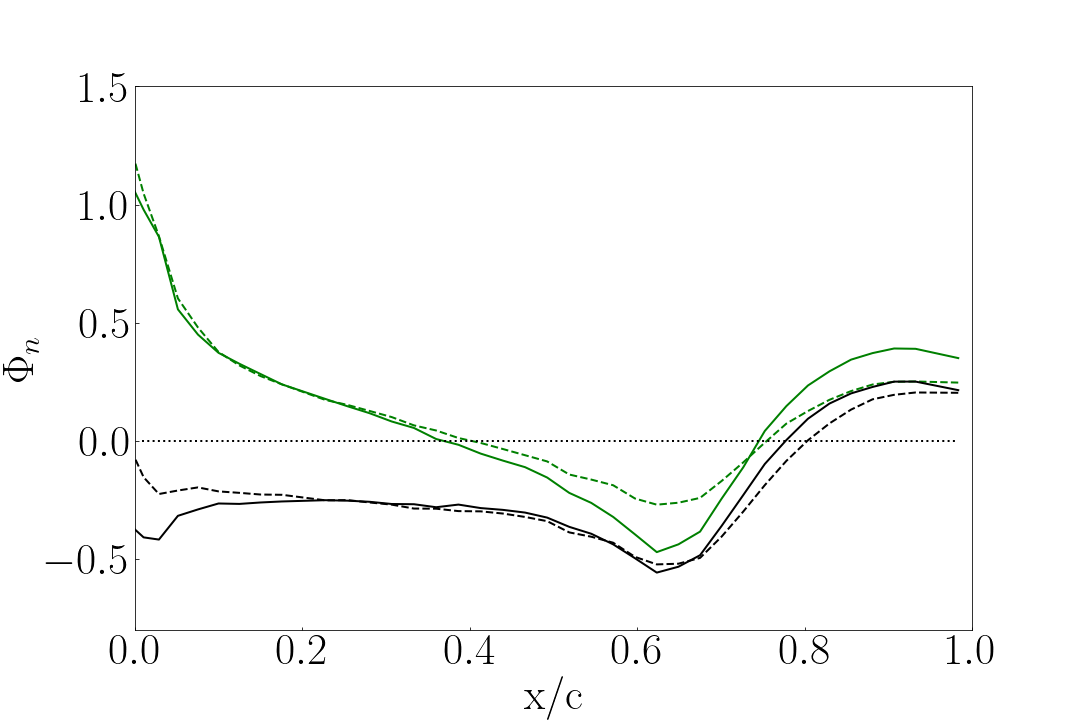} & \includegraphics[trim=2.5cm 0 3.cm 0, clip,height=4.5cm]{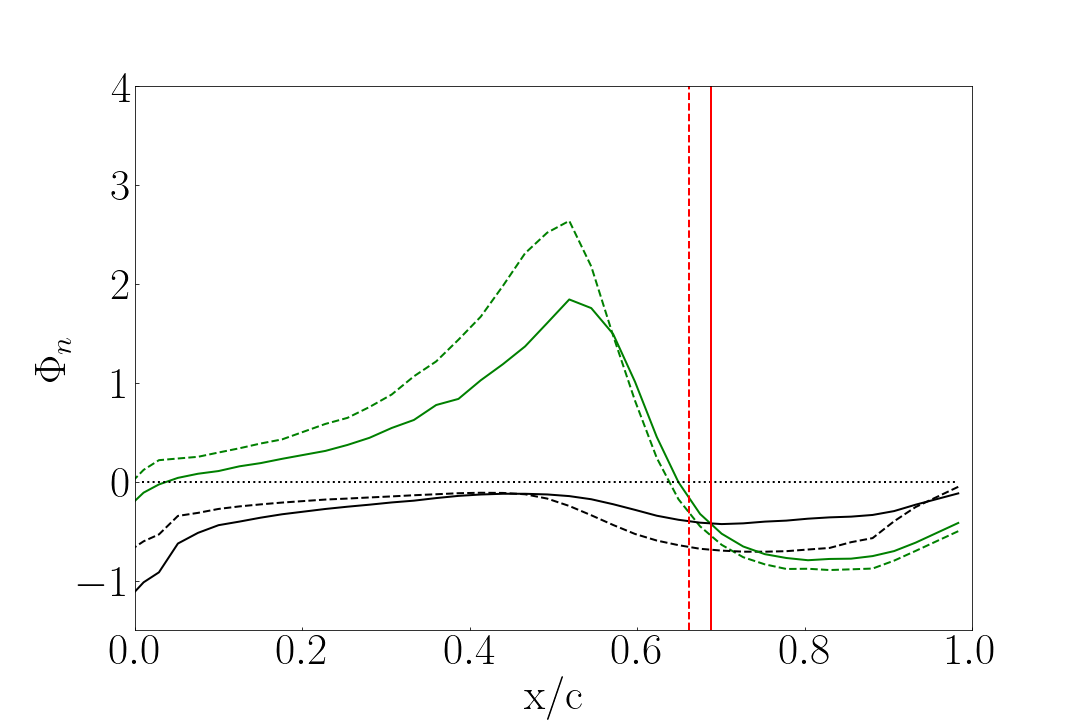} & \includegraphics[trim=2.5cm 0 3.cm 0, clip, height=4.5cm]{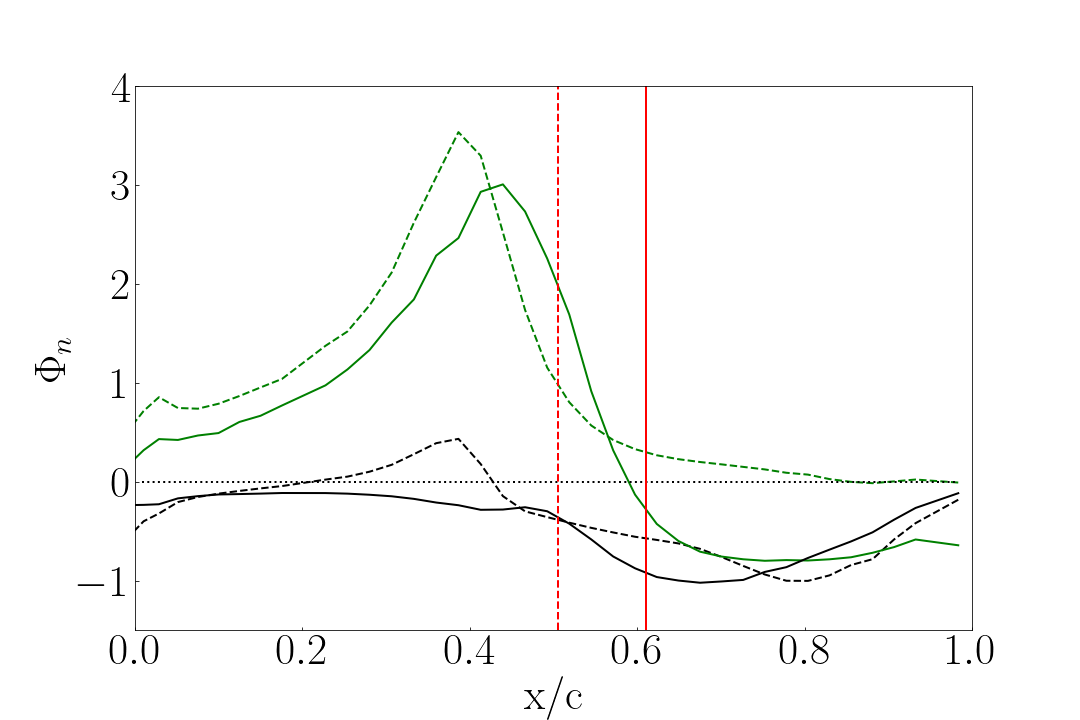} \\ 
   \small{$AoA=16 ^\circ$}  & \small{$AoA=20 ^\circ$ } & \small{$AoA=24 ^\circ$ } \\
 \includegraphics[trim=2.5cm 0 3.cm 0, clip,height=4.5cm]{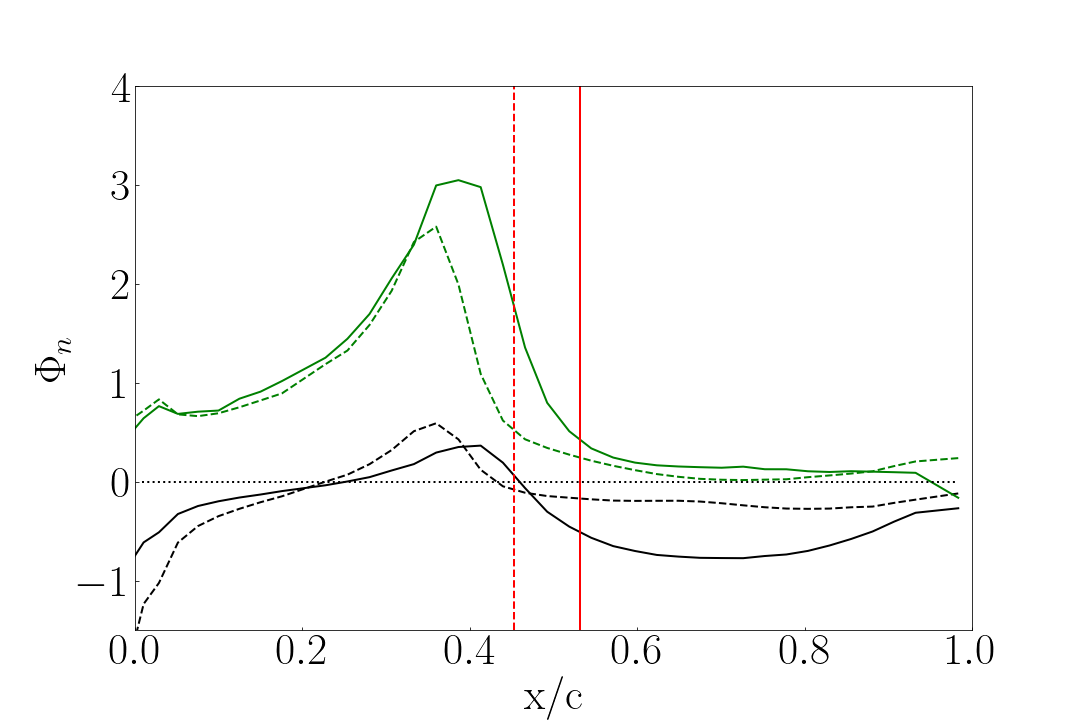} & \includegraphics[trim=2.5cm 0 3.cm 0, clip,height=4.5cm]{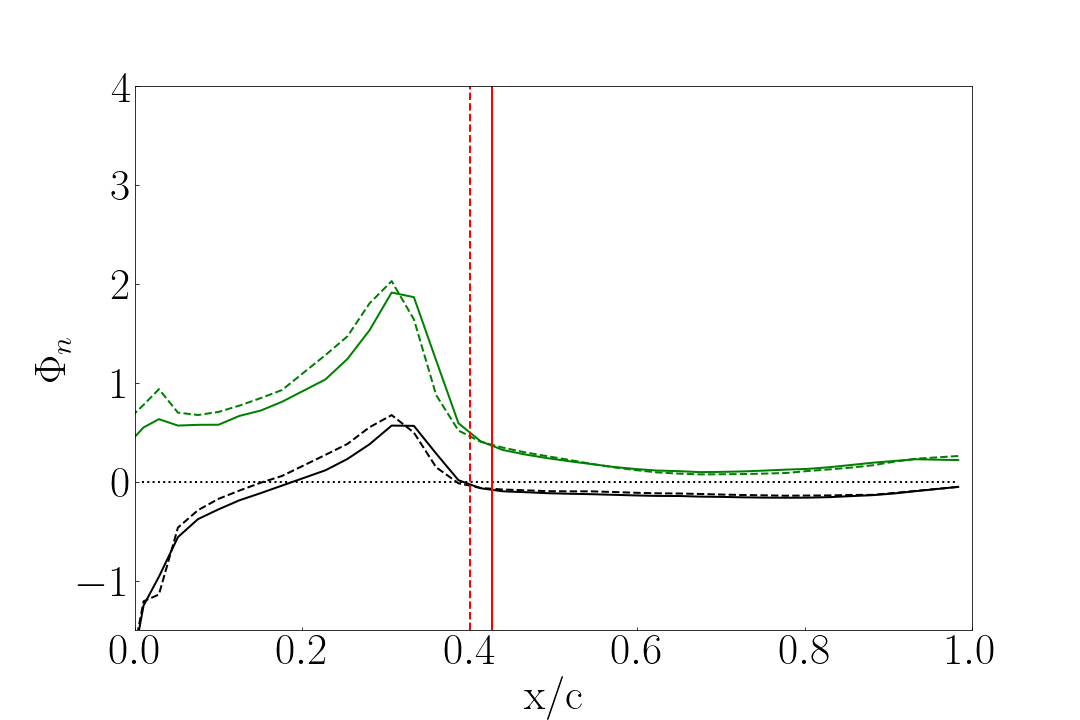} & \includegraphics[trim=2.5cm 0 3.cm 0, clip, height=4.5cm]{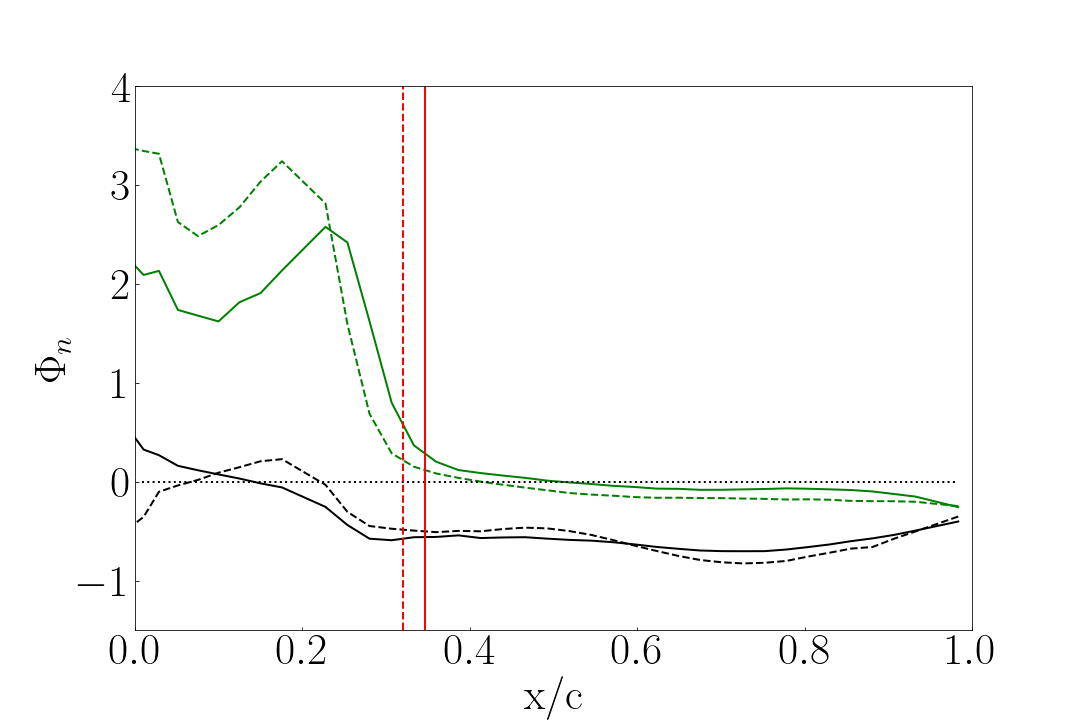} \\ 
\end{tabular}
\renewcommand{\arraystretch}{1}
\caption{First two POD modes of the pressure coefficient $\Phi_n \lambda_n^{1/2}$ on the suction side for $Y^-$ (solid lines) and $Y^+$ (dashed lines). The red solid and dashed lines respectively correspond to the steady separation point at $Y^+$ and $Y^-$. } 
\label{BERENGERE/modesingle}
\end{figure}

Figure \ref{BERENGERE/modesingle} represents the modes $\Phi_n$ scaled with their average contribution $\lambda_n^{1/2}$. 
At low angles of attack ($\le$ \ang{8}), the shape of the modes remains similar with a small peak in the leading edge region and then a slow decrease along the chord.
No significant differences are observed between the two spanwise locations, $Y^{+}$ and $Y^{-}$, which confirms that the flow is statistically 2D for low angles of attack. 
Just before flow separation at \ang{10},  the first two modes look very similar over the aft region of the suction side. 
Both modes are characterized by a local maximum corresponding to the maximum variance location aka the intermittent separation point.
At \ang{12}, the  shapes of the modes change drastically and will remain more or less similar at higher angles.
We can see that the sharp changes in POD energy levels displayed in figure \ref{lambdan12} at \ang{12} correspond to changes in the spatial structures of the modes. 
The vertical lines in the figure correspond to the steady separation criterion, applied to the time-averaged pressure coefficient, as defined in section~\ref{sec:separation_point}. 
One can see that the steady separation point constitute the downstream limit of the region of strong fluctuations similarly as observed in section \ref{subsec:pressureDist} from the pressure gradient and fluctuation distribution. It follows closely the location of the dominant mode maximum which is by definition that of the intermittent separation point.
We note that the separation region at $Y_+$ is located slightly farther upstream than that at $Y_-$, with largest discrepancies at \ang{14} and \ang{16}. This indicates again the loss of 2D statistical representation of the flow separation region.
The shape of the modes remains essentially similar up to \ang{24}, when the region of strong adverse pressure gradient merges with the leading edge.

\begin{figure}[htbp]
\centerline{\includegraphics[ trim=0cm 0 0cm 0, clip, height=2cm]{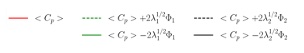}}
\begin{tabular}{ccc}
\includegraphics[ trim=0cm 0 0cm 0, clip,height=4cm]{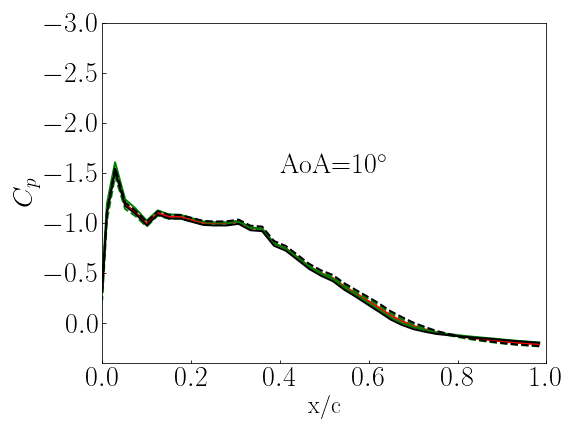} &
\includegraphics[trim=0cm 0 0cm 0, clip,height=4cm]{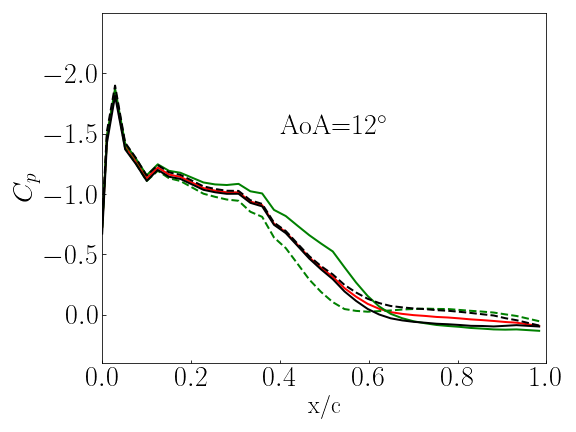} &
 \includegraphics[trim=0cm 0 0cm 0, clip, height=4cm]{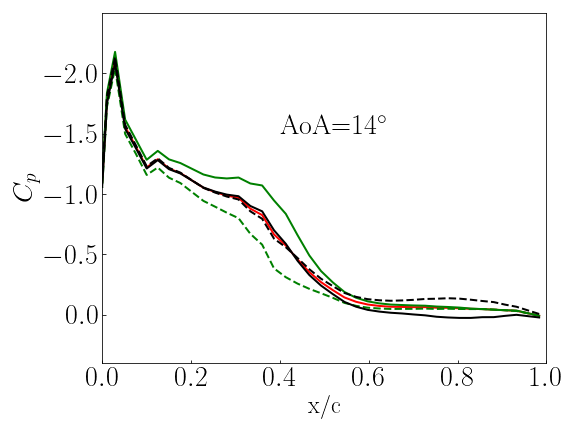} \\
 \includegraphics[ trim=0cm 0 0cm 0, clip,height=4cm]{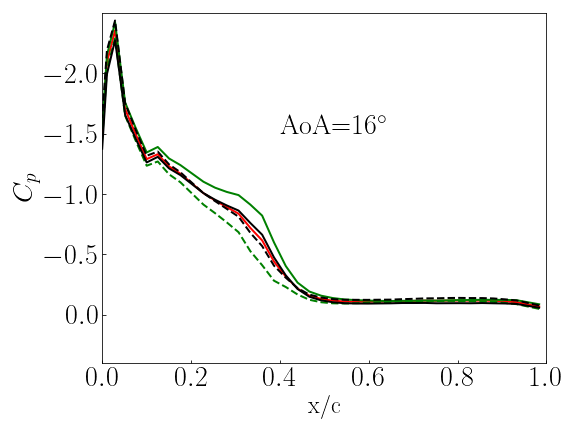} &
  \includegraphics[trim=0cm 0 0cm 0, clip,height=4cm]{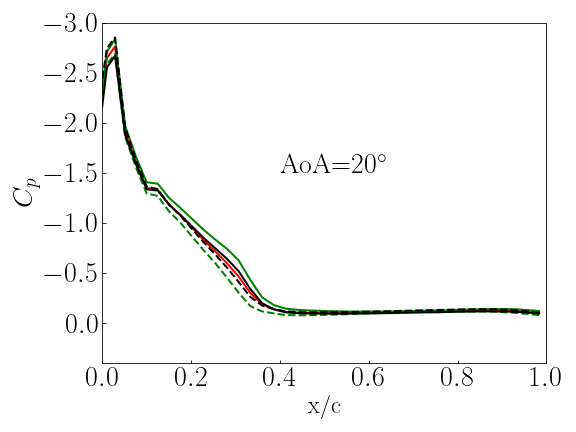} &
 \includegraphics[trim=0cm 0 0cm 0, clip, height=4cm]{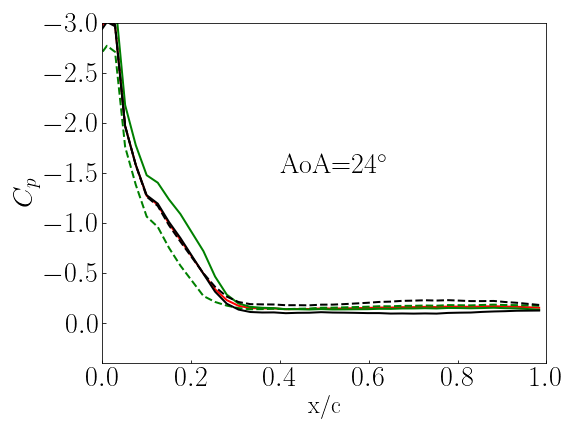} \\
\end{tabular}
\caption{ POD-based reconstructions of the pressure coefficient on the suction side for different angles of attack using the two first modes (the modes represented are those determined at $Y_+$ - the same trends  were observed for those obtained at $Y_-$).}
\label{fig:cppod}
\end{figure}

\berengere{ To determine the contribution of each mode to the spatial distribution of the pressure coefficient , we represented in figure \ref{fig:cppod} a POD-based reconstruction of the pressure coefficient using the first two modes and selected values of the mode amplitudes.
We can see that when the amplitude of mode 1 is negative (resp. positive), the reconstructed pressure coefficient curve has a shorter plateau (resp. longer plateau) at zero in the trailing edge region, a more intense (resp. less intense) adverse pressure gradient region, thus corresponding to a separation closer to the trailing edge (resp. closer to the leading edge).
As expected, no significant contribution from the modes is observed for low angles of attack smaller than (and including) \ang{10}. 
At the angle of attack \ang{12}, the reconstruction based on the dominant mode $\Phi_1$ shows a modification of both the trailing edge region (fluctuation of the local maximum) and the recovery region (displacement of the  separation point). 
In contrast, the second mode only contributes to the fluctuations in the separated region close to the trailing edge, with a contribution level that is equivalent to that of mode 1.
Since the fluctuations in the recovery region are mostly due to mode 1, and those at the trailing edge are about equally due to mode 1 and mode 2 (which are by construction uncorrelated), this means that the recovery region and the trailing edge region are partially correlated. 
In contrast, figure \ref{fig:cppod} shows that at \ang{14}, the fluctuations in the region of adverse pressure gradient, in the intermittent separation region, are completely decorrelated from those in the separated region over the trailing edge, as the fluctuations in the region of adverse pressure gradient is exclusively carried by mode 1 and the separated region by mode 2.
Figure \ref{fig:cppod} shows at \ang{16} that the intensity of the fluctuations over the trailing edge subsides, which is consistent with the displacement of the separated shear layer away from the airfoil.
For the angle of attack of \ang{20}, 
mode 1 creates less intense, but still significant fluctuations in the adverse pressure gradient region, but both mode 1 and mode 2 are also associated with fluctuations at the leading edge at \ang{20}.
For the highest angle of attack \ang{24}, mode 1 is again dominant over a region that extends from the leading edge to
the separation point.}
Overall, these observations support the idea that the onset of flow separation is an inherently three-dimensional, continuously evolving process.

\berengere{ 
To sum up, the pressure fluctuations on each chord are mostly captured by the first two POD modes.
At angles larger than \ang{10}, the first mode represents the variations of the  pressure gradient and is dominant in the intermittent separation region.
For low angles of attack (\ang{10} and \ang{12}), both mode 1 and mode 2 contribute to the flow separation near the trailing edge, but 
at larger angles ($AoA \geq \ang{14}$), as the separation point moves towards the leading edge,
mode 1 becomes decorrelated from the separated flow at the trailing edge. }

\subsection{ POD characterization of the bi-stability }

\begin{figure}[htbp]
\begin{tabular}{cccc}

\includegraphics[ trim=0.0cm 0 0.0cm 0, clip,height=4cm]{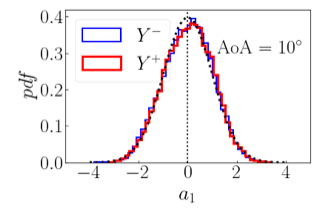} & \includegraphics[trim=0.0cm 0 0.0cm 0, clip,height=4cm]{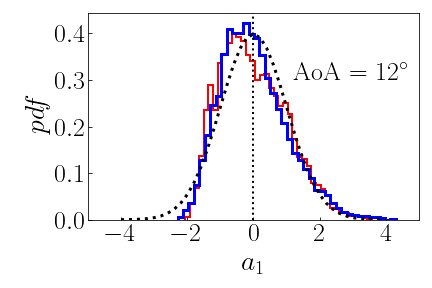} & \includegraphics[trim=0.0cm 0 0.0cm 0, clip, height=4cm]{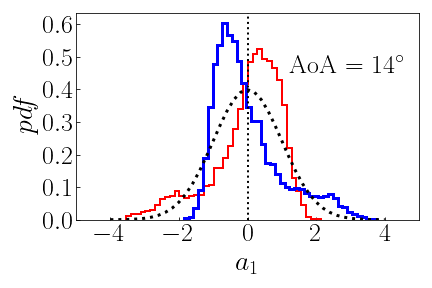} \\

\includegraphics[trim=0.0cm 0 0.0cm 0, clip,height=4cm]{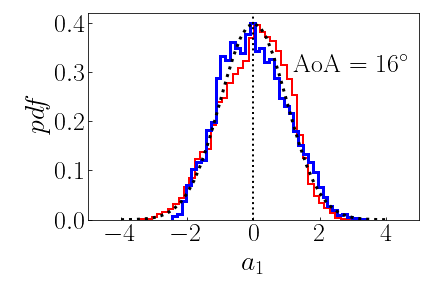} & \includegraphics[trim=0.0cm 0 0.0cm 0, clip,height=4cm]{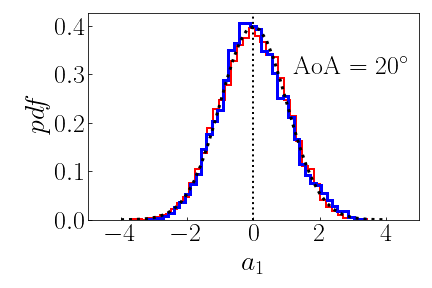} & \includegraphics[trim=0.0cm 0 0.0cm 0, clip, height=4cm]{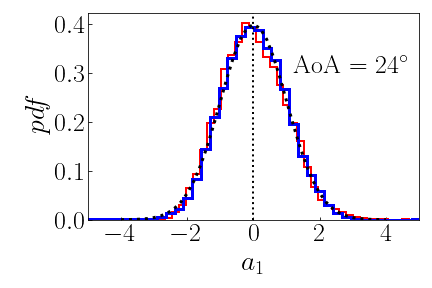} \\
\end{tabular}
\caption{ Histogram of the first two POD amplitudes - the black dotted line indicates a reference Gaussian distribution.}
\label{histcoef1}
\end{figure}

\begin{figure}[htbp]
\begin{tabular}{ccc}
\includegraphics[ trim=0cm 0 0cm 0, clip,height=4cm]{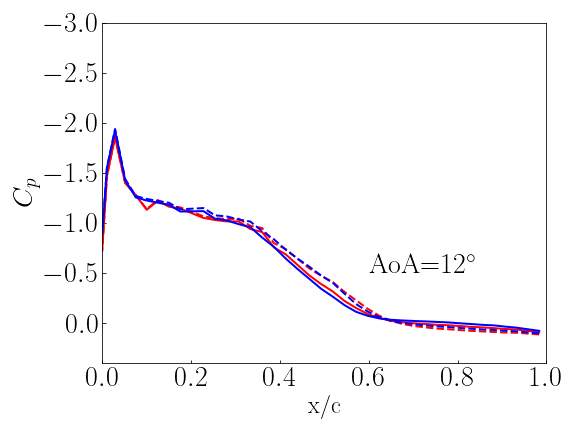} & 
\includegraphics[ trim=0cm 0 0cm 0, clip,height=4cm]{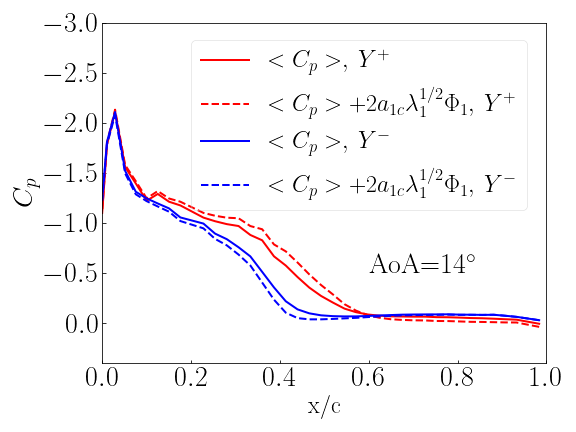} & \includegraphics[trim=0cm 0 0cm 0, clip,height=4cm]{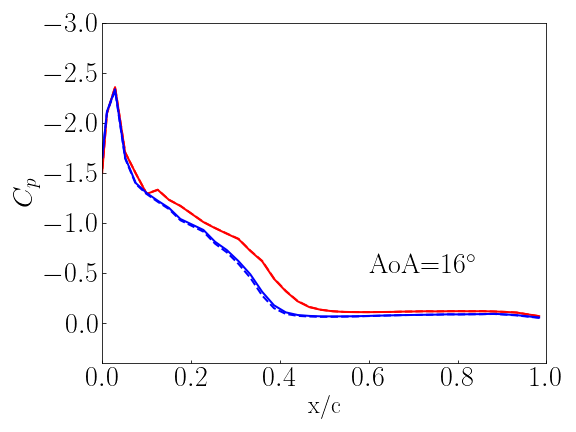} \\
\end{tabular}
\caption{POD-based reconstructions of the pressure coefficient on the suction side for different angles of attack using the dominant mode
and the most frequent value of the amplitude (a rescaling factor of 2 has been applied to the fluctuation for easier visualization). }
\label{fig:cpchar}
\end{figure}

We now focus on the temporal amplitude of the dominant mode $a_1(t)$ for both chords. This amplitude  is almost entirely correlated with the pressure signal at the ISP (with a correlation coefficient larger than 0.95).
A distribution of the different values taken by the normalized amplitude $a_1(t)$ in time can be represented as a histogram (figure \ref{histcoef1}), which provides a description of the dominant mode dynamics.
In particular, the mode (most frequent value) of the histogram $a_{1c}$ corresponds to the most likely configuration of the flow, which may be different from zero (its time-averaged value) if the distribution is asymmetric.
The following observations did not change when half the time period was used, as well as when we considered longer datasets obtained at \ang{14} and \ang{16}.
At the angle of attack of \ang{10} (and for lower angles as well, not shown), the distribution of the dominant POD amplitude is nearly symmetric and close to a Gaussian.
Deviations from Gaussianity are observed at \ang{12}. 
The distributions have the same asymmetry, with more frequent negative values, which means that the time-averaged separation point is on average located downstream of its most frequent position.  
This is illustrated in figure \ref{fig:cpchar}, which represents for each chord the mean pressure coefficient $<C_p>$ and the reconstructed pressure based on the dominant fluctuation $<C_p>+ 2 a_{1c} \phi_1$, where $a_{1c}$ is the most likely value of the amplitude and the rescaling of 2 was chosen to make visualization easier. This means that the separation region makes large, infrequent excursions upstream on both chords.

At an angle of attack of \ang{14}, corresponding to the maximum level of fluctuations, figure \ref{histcoef1} shows that the deviation from Gaussianity is maximal. The remarkable observation here is that the distributions at $Y^+$ and $Y^-$ are now antisymmetric, with a high negative (resp. positive) tail for $Y^-$ (resp. $Y^+$). 
This means that the separation region at $Y^-$ makes large, infrequent excursions towards the leading edge (like both chords at \ang{12}), while the large, infrequent excursions at $Y^+$ are now directed towards the trailing edge. 
As can be seen in figure \ref{fig:cpchar}, this results in  a persistent dissymmetry between the two chords, so that 
the separation point at $Y^+$ is located closer to the trailing edge with a more intense pressure gradient than that at $Y^-$.  
Figure \ref{histcoef1} shows that the antisymmetry of the distributions is much weakened at \ang{16}, while the time-averaged dissymmetry (see figure \ref{fig:cpchar}) is still significant.
For angles of attack larger than \ang{20}, the distribution of the amplitude of the two dominant POD modes appears to be Gaussian again.

We have established that the amplitude of the dominant POD mode captures the dynamics of the most important flow phenomenon for each angle of attack, namely the evolution of the separation point and the flow separation for higher angles of attack.
A characterization of bi-stability can further be provided by the correlation of the dominant POD amplitudes.

Let $a_n^+(^-)$ be the amplitude of mode $n$ at the spanwise locations $Y^+$ ($Y^-$).
Table \ref{correl} represents the correlation coefficient between the amplitudes of the first two POD modes $C(a_n^-,a_n^+)$ for different angles of attack. 
For the lowest angles ($AoA \le$ \ang{8}), a strong positive correlation is observed for the first two modes, indicating strong spatial coherence all over the suction side. 
At \ang{10}, when the flow starts to detach, evidence of correlation disappears. 
At \ang{12}, when the flow separation starts to move upstream, the amplitudes of the dominant mode become negatively correlated.
The negative correlation is maximal (-0.7) for the angles of attack \ang{14} and \ang{16}, then disappears at \ang{20}.
At \ang{24}, a positive correlation is observed with the dominant mode, which is now strongest over the leading edge.
The correlation coefficient therefore varies (in absolute value)  like the normal force standard deviation shown in figure~\ref{fig:CN}.
This means that in the bistability regime, $C(a_n^-,a_n^+)$  constitutes a good indicator of the local switch phenomenon described above.
It also shows that local switches represent an essential part of the pressure dynamics over the complete airfoil section. 
Bi-stability therefore appears as a characteristic of early separation, before complete flow separation over the airfoil.

\begin{figure}[htbp]
\begin{tabular}{c}
\includegraphics[trim=0cm 0 0cm 0, clip,height=8cm]{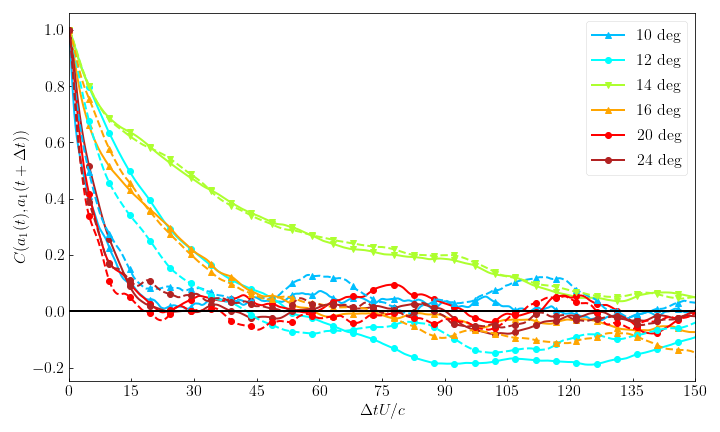} \\
\includegraphics[trim=0cm 0 0cm 0, clip,height=8cm]{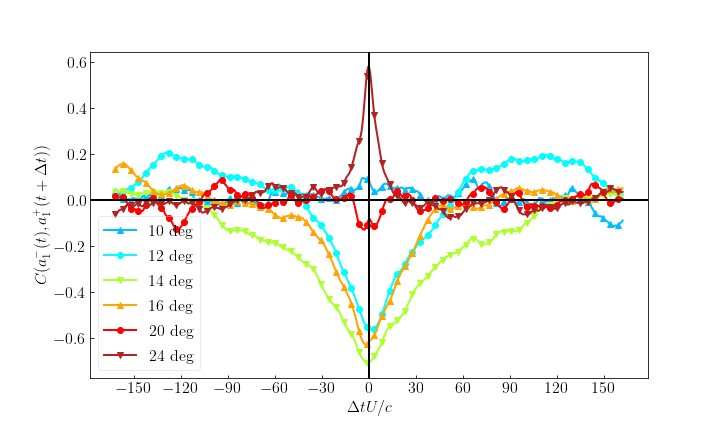} \\
\end{tabular}
\caption{
Autocorrelation of the dominant (normalized) POD amplitude for different angles of attack;   
Cross-correlation of  the dominant  POD amplitude for different angles of attack .} 
\label{fig:autocorrelation}
\end{figure}

\berengere{ Global time scales can be extracted from the autocorrelation and the cross-correlation of the POD amplitudes, shown in figure
~\ref{fig:autocorrelation}, which can be respectively connected with the local jumps and switches described above.
Examination of the autocorrelation of $a_1$ (figure \ref{fig:autocorrelation} a)) confirms bi-stability is not a cyclical process as the cross-correlation falls to zero. 
The time delay at the first zero crossing $T_0$  gives a measure of the time during which the signal remains correlated with itself. It can be seen
that this time scale is both on the order of and follows the variations of the local switch time scale $T$ with the angle of attack. 
$T_0$ is on the order of 15 convective time scales for \ang{10}, then increases significantly to about 40 for \ang{12} and to a maximum of 150 at \ang{14}.
For higher angles, $T_0$  decreases to about 45  for \ang{16} and to 15  for $AoA \ge \ang{20}$.
A similar  time scale is present in the cross-correlation of the dominant POD amplitude in the [\ang{12}, \ang{20}] 
regime, which is represented in figure \ref{fig:autocorrelation}b). 
We note that the modulus of the cross-correlation is close to its maximum value at zero time delay, which suggests that switches 
are essentially  synchronized.}

\renewcommand{\arraystretch}{1.0}
\begin{table}

\begin{tabular}{|l|c|c|c|c|c|c|c|c|c|}
\hline
Angle of attack & 0 & 4 & 8 & 10 & 12 & 14 & 16 & 20 & 24 \\ \hline
Mode 1 & 
 {\bf 0.81} & 
  {\bf 0.52} & 
  {\bf 0.40} & 
  0.018 & 
 {\bf -0.57} &
  {\bf -0.71} & 
 {\bf -0.62} &
 -0.086 &
 {\bf 0.57} \\ \hline
 Mode 2  &
 0.15  
&{\bf 0.38}  
&{\bf 0.47}  
&0.27  
&-0.28 
&0.056 
&0.10 
&0.15 
&  0.30  \\ \hline

\end{tabular}
\caption{Correlation coefficients between the amplitudes of the first two POD modes at the two spanwise locations $Y^+$ and $Y^-$. Coefficients larger than 0.3 in absolute value are indicated in boldface typescript.}
\label{correl}
\end{table}

\section{Conclusions}
\label{sec:conclusions}

We have investigated the behavior of wall pressure measurements over a moderately thick (20\% of the chord) and cambered (4\% of the chord) airfoil at a high chord-based Reynolds number, $Re_c=$ \num{3.6e6}, and over a range of angles of attack including the maximum force.
The airfoil was equipped with synchronized unsteady wall pressure sensors distributed along the chord at two spanwise locations, symmetrically from the mid-span of the blade.
At angles of attack where the normal force reaches its maximum, the load fluctuations are also the largest which can cause additional fatigue and damage on wind turbines.
\berengere{ Evidence of asymmetry between the chords was found in the time-averaged statistics in a range of angles of attack $[\ang{12}, \ang{20}]$.
In this asymmetric regime, the largest fluctuations were concentrated in the region of strong adverse pressure gradient.} 
The location of the maximum of the fluctuations moved towards the leading edge with the increase of the angle of attack 
and was highly correlated with the location of the pressure distribution inflection point. 
 The location of the maximum of the fluctuations therefore appeared as a relevant indicator of the intermittent separation region. 
\berengere{ We therefore called it the intermittent separation point (ISP).}

\berengere{ This regime was also characterized by  a bi-stability phenomenon:}
the wall pressure signals at the maximum of pressure fluctuation displayed a jump-like character between two characteristic levels, which 
were shown to correspond to intermittent flow separation and reattachment.
In addition, the jumps on each chord were highly anti-correlated, with an anticorrelation maximum between the two fluctuation peak locations.
The physical description of the bi-stability phenomenon can be given as follows: when the intermittent separation region on the chord $Y^+$ (resp. $Y^-$) moves downstream, thereby increasing the area of attached flow on that side, the intermittent separation region on the other chord $Y^-$ (resp. $Y^+$) moves upstream, which yields a larger separated flow area.
A characteristic time scale based on the switches measured at the ISP was proposed to characterize the bistability phenomenon.
\berengere{ The average time between  switches was found to evolve with the variance of the pressure fluctuations.}

POD analysis provides additional insight into this description, as a large part of the fluctuation energy (over 70 \%) could be reconstructed with only two spatial modes.
The amplitude of the dominant mode displays a very high correlation (larger than 0.9) with the intermittent separation point, which confirms the relevance of this location.
Independent two-mode reconstructions of the instantaneous signal at the two spanwise locations are able to capture the highly intermittent flow separation and reattachment phenomena linked with pressure jumps at the intermittent separation point. 
The reconstructions also showed that as the angle of attack increases, the region of maximum fluctuations associated with the strong adverse pressure gradient became progressively decorrelated from the trailing edge and moves towards the leading edge, eventually merging with it.
The evolution of the flow as it transitioned into bi-stability could be described by a single indicator based on the correlation coefficient between the dominant POD mode amplitudes on each chord.
Overall, our results indicate that flow separation at high Reynolds numbers is an inherently local, three-dimensional and unsteady process that occurs in a continuous manner and leads to high load fluctuations when the maximum of aerodynamic force is generated. 
However, as the characteristics of the flow separation can be represented with mainly 2 POD modes, our results also suggest that a low-order approach may offer a viable route to modelling and ultimately predicting and controlling this complex bi-stability phenomenon.

\begin{acknowledgements}
The experiments were performed within the research projects ePARADISE with the funding from ADEME and Pays-de-Loire region in France (grant no. 1905C0030). 
Analyses were finalized within the French-Swiss project MISTERY funded by the French National Research Agency (ANR PRCI grant no. 266157) and the Swiss National Science Foundation (grant no. 200021L 21271). 
\berengere{The authors are thankful to the anonymous reviewers for their helpful suggestions.} 
\end{acknowledgements}